\documentclass[12pt]{article}

\usepackage[dvipsnames]{xcolor}
\usepackage{graphicx}
\usepackage{amsmath}        
\usepackage{amssymb}        
\usepackage{slashed}
\usepackage{bm}
\usepackage{here}
\usepackage{cite}

\def\mydate{October 26,  2026}
\def\ignore#1{{}}

\def\go{\rightarrow}
\def\dd{\partial}

\def\ep{{\epsilon}}

\def\SM{{\rm SM}}
\def\KK{{\rm KK}}
\def\EM{{\rm EM}}

\def\max{{\rm max}}
\def\onehalf{\hbox{$\frac{1}{2}$}}

\def\la{\langle}
\def\ra{\rangle}

\def\Tr{{\rm Tr} \,}

\def\cD{{\cal D}}

\def\mybig{\displaystyle \strut }

\def\myfrac#1#2{\frac{\mybig #1}{\mybig #2}}

\def\mymat#1#2{\begin{matrix}#1 \cr \noalign{\kern -2pt} #2\end{matrix}}

\def\mynoalign{\noalign{\kern 4pt}}
\def\mysnoalign{\noalign{\kern 3pt}}
\def\mytinynoalign{\noalign{\kern 2pt}}
\def\ignore#1{{}}

\makeatletter
\@addtoreset{equation}{section}
\makeatother

\begin{document}

\thispagestyle{empty}

{\small \noindent \mydate   \hfill }

\vskip 2.5cm

\baselineskip=35pt plus 1pt minus 1pt

\begin{center}
{\bf \LARGE Anomaly in baryon number}\\ 

\end{center}


\baselineskip=22pt plus 1pt minus 1pt

\vskip 1.5cm

\begin{center}
\renewcommand{\thefootnote}{\fnsymbol{footnote}}
{\bf  Yutaka Hosotani\footnote[1]{hosotani@rcnp.osaka-u.ac.jp}}

\baselineskip=18pt plus 1pt minus 1pt

\vskip 10pt
{\small \it Research Center for Nuclear Physics, University of Osaka,}\\
{\small \it Ibaraki, Osaka 567-0047, Japan}\\

\end{center}

\vskip 2.cm
\baselineskip=18pt plus 1pt minus 1pt

\begin{abstract}
Chiral anomalies in baryon and lepton number currents in  the GUT-inspired $SO(5) \times U(1) \times SU(3)$  
gauge-Higgs unification model in the  Randall-Sundrum (RS)  warped  space are examined.
Total anomalies including contributions of all Kaluza-Klein (KK) excited modes of fermions running along 
internal triangular loops are expressed  in terms of the values of  wave functions  of gauge-boson 
KK towers at the ultraviolet (UV) and infrared (IR) branes in the RS space. 
The covariant divergence of 5D baryon number current 
picks up an anomaly term proportional to 
$\Tr F_{\mu\nu} \tilde F^{\mu\nu} |_{SU(2)_L} - \Tr F_{\mu\nu} \tilde F^{\mu\nu} |_{SU(2)_R}$
at the UV and IR branes where $SU(2)_L \times SU(2)_R$ is a subgroup of $SO(5)$.

\end{abstract}

\newpage

\baselineskip=20pt plus 1pt minus 1pt
\parskip=0pt

\section{Introduction} 

In the standard model (SM) of $SU(2)_L \times U(1)_Y \times SU(3)_C$ gauge theory, 
baryon number is conserved at the classical level, but the baryon number symmetry
suffers from anomalies at the quantum 
level\cite{Adler1969, BellJackiw1969, Bardeen1969, Fujikawa1979, Manton1984, Donoghue2014, Smith2026}. 
The baryon number current $j_B^\mu (x)$ is not conserved, but
\begin{align}
&\dd_\mu j_B^\mu = - \frac{3}{32 \pi^2}  \, 
\Big\{ g_w^2 \sum_{a=1}^3 F^a_{\mu\nu} \tilde F^{\mu\nu \, a} \big|_{SU(2)_L}  
- g^{\prime 2} B_{\mu\nu} \tilde B^{\mu\nu}  \big|_{U(1)_Y} \Big\} ~,   
\label{SManomaly1}
\end{align}
where 
$g_w$ and $g'$ are the $SU(2)_L$ and $U(1)_Y$ gauge couplings, respectively. 
The anomaly terms in Eq.\ (\ref{SManomaly1}) arise from triangle diagrams in which
quarks run along internal loops.  The anomaly term could play an important role in the process
of baryon number generation in the early universe.
Although the SM has been successful in describing  phenomena at low energies, it has a gauge hierarchy problem.  
Gauge-Higgs unification (GHU) models have been proposed to solve the gauge hierarchy problem in which
the 4D Higgs boson is identified with a part of  gauge fields in five dimensions \cite{Hosotani1983, Davies1988,
Hosotani1989, Davies1989, Hatanaka1998, Kubo2002, BurdmanNomura2003, Csaki2003, Scrucca2003, 
 GUTinspired2019a, YHbook}.
Then a question arises about how the anomaly equation (\ref{SManomaly1}) in the SM is modified  
in five-dimensional gauge theories.
Years ago Arkani-Hamed, Cohen and Georgi examined a chiral $U(1)$ gauge theory on $M^4 \times (S^1/Z_2)$ orbifold
and showed that chiral anomalies appear on the orbifold fixed planes \cite{Arkani-Hamed2001}.
We generalize their argument to realistic GHU models.

Among various GHU models the $SO(5)\times U(1) \times SU(3)$ GHU in the Randall-Sundrum (RS) 
warped  space is a promising realistic model\cite{GUTinspired2019a}.
It is inspired by $SO(11)$ gauge-Higgs grand unification models \cite{SO11GHGU}, 
and  the GUT-inspired GHU model gives nearly the same phenomenology at low energies as in the SM.  
The Cabibbo-Kobayashi-Maskawa  (CKM) matrix in the quark sector is  reproduced \cite{CKM2025}, 
and neutrino oscillations are explained with  the Pontecorvo-Maki-Nakagawa-Sakata (PMNS) 
matrix  in the normal ordering\cite{neutrino2026}.
It also predicts distinct deviation from the SM in forward-backward asymmetry in fermion pair production 
in electron-positron ($e^- e^+$)  collisions  and smaller Higgs cubic and quartic self-couplings than those 
in the SM\cite{YHbook}.
In the present paper we derive an anomaly formula for the baryon number current 
in  the GUT-inspired GHU model.

In the previous papers anomaly flow and cancellation of gauge anomalies in the GUT-inpired GHU model 
were investigated \cite{anomaly2026, AnomalyFlow1, AnomalyFlow2}.
It was shown there that  the total gauge anomalies including contributions of all KK excited modes of 
fermions become universal, and cancellation of gauge anomalies is achieved in each generation.
The technique developed there can be applied to global anomalies such as baryon number anomaly.
We show that after incorporating contributions of all KK excited modes of quark multiplets 
anomaly terms in baryon number currents are  proportional to 
$\Tr F_{\mu\nu} \tilde F^{\mu\nu} |_{SU(2)_L} - \Tr F_{\mu\nu} \tilde F^{\mu\nu} |_{SU(2)_R}$
at the UV and IR branes where $SU(2)_L \times SU(2)_R$ is a subgroup of $SO(5)$.

In Section 2 the GUT-inspired GHU model is described. 
In Section 3 baryon and lepton number currents in both five and four dimensions are defined. 
In Section 4 chiral anomalies of all 4D gauge fields are evaluated, and in Section 5 a 5D anomaly formula 
is derived.  A brief summary is given in Section 6.
In appendix A basis functions in the RS space are given.
In appendix B wave functions of gauge fields are given.
In appendix C numerical results for various $F$ factors, which become important in the anomaly formulas, are given.


\section{GUT-inspired GHU model} 
 
 We examine baryon number current in the GUT-inspired $SO(5) \times U(1)_X \times SU(3)_C (\equiv {\cal G})$ 
GHU  \cite{GUTinspired2019a, YHbook}.  
The model is defined in the RS warped space with the metric \cite{RS1}
\begin{align}
ds^2= G_{MN} \,  dx^M dx^N =
e^{-2\sigma(y)} \eta_{\mu\nu}dx^\mu dx^\nu+dy^2,
\label{RSmetric1}
\end{align}
where $M,N=0,1,2,3,5$, $\mu,\nu=0,1,2,3$, $y=x^5$, $\eta_{\mu\nu}=\mbox{diag}(-1,+1,+1,+1)$,
$\sigma(y)=\sigma(y+ 2L)=\sigma(-y)$, and $\sigma(y)=ky$ for $0 \le y \le L$.
The RS space has the same spacetime topology as $M^4 \times (S^1/Z_2)$.  
In terms of the conformal coordinate $z=e^{ky}$ ($0 \le y \le L$, $1\leq z\leq z_L=e^{kL}$)
the metric becomes
\begin{align}
ds^2=  \frac{1}{z^2} \bigg(\eta_{\mu\nu}dx^{\mu} dx^{\nu} + \frac{dz^2}{k^2}\bigg)~ .
\label{RSmetric-2}
\end{align}
The bulk region $0<y<L$ is anti-de Sitter spacetime 
with a cosmological constant $\Lambda=-6k^2$.  It is sandwiched by the
UV brane at $y=0$  and the IR brane at $y=L$.  
The KK mass scale is given by $m_{\rm KK}=\pi k/(z_L-1) \simeq \pi kz_L^{-1}$.
Typical values are $z_L \sim 10^{11}$  and $m_\KK  \sim 13\,$TeV.

Gauge fields $A_M^{SO(5)}$,  $A_M^{U(1)_X}$ and $A_M^{SU(3)_C}$ of 
$SO(5) \times U(1)_X \times SU(3)_C$ with gauge couplings  $g_A$,  $g_B$ and $g_C$
satisfy the orbifold boundary conditions (BCs)
\begin{align}
\begin{pmatrix} A_\mu \cr  A_{y} \end{pmatrix} (x,y_j-y) &=
P_{j} \begin{pmatrix} A_\mu \cr  - A_{y} \end{pmatrix} (x,y_j+y)P_{j}^{-1}
\quad (j=0,1), \cr
\noalign{\kern 5pt}
(y_0, y_1) &= (0, L) .
\label{BC-gauge1}
\end{align}
Here   $P_0=P_1 = P_{\bf 5}^{SO(5)} =\mbox{diag} (I_{4},-I_{1} )$ for $A_M^{SO(5)}$ in the vector 
representation and  $P_0=P_1 = P_{\bf 4}^{SO(5)} =\mbox{diag} (I_{2},-I_{2} )$  for 
$A_M^{SO(5)}$ in the spinorial  representation.    $P_0=P_1= 1$ for $A_M^{U(1)_X}$ and $A_M^{SU(3)_C}$.
The orbifold BCs break $SO(5)$ to $SO(4) \simeq SU(2)_L \times SU(2)_R$.
In the following we write $A_M^{SO(5)} = A_M$ and  $A_M^{U(1)_X} = B_M$ unless confusion arises.

The 4D Higgs boson $\phi_H(x)$ is  a zero-mode  contained in  $A_z = (kz)^{-1} A_y$;
\begin{align}
A_z^{(j5)} (x, z) &= \frac{1}{\sqrt{k}} \, \phi_j (x) u_H (z) + \cdots , ~~
u_H (z) = \sqrt{ \frac{2}{z_L^2 -1} } \, z ~, \cr
\noalign{\kern 5pt}
\phi_H(x) &= \frac{1}{\sqrt{2}} \begin{pmatrix} \phi_2 + i \phi_1 \cr \phi_4 - i\phi_3 \end{pmatrix} .
\label{4dHiggs}
\end{align}
We take $\la \phi_1 \ra , \la \phi_2 \ra , \la \phi_3 \ra  =0$ and   $\la \phi_4 \ra \not= 0$
without loss of generality.  
The AB phase $\theta_H$ in the fifth dimension is given by $\la \phi_4 \ra  = \theta_H f_H$ where
\begin{align}
&P \exp \bigg\{ i g_A \oint dy \, A_y \bigg\}
\sim \exp \bigg\{ i \Big(  \theta_H + \frac{H(x)}{f_H} \Big) \, 2 \, T^{(45)}\bigg\} ~, \cr
\noalign{\kern 5pt}
&f_H  = \frac{2}{g_w} \sqrt{ \frac{k}{L(z_L^2 -1)}} ~~,~ g_w = \frac{g_A}{\sqrt{L}} ~.
\label{fH1}
\end{align}
Physics is periodic in $\theta_H$ with a period $2 \pi$.  
The neutral Higgs field $H(x)$ is a four-dimensional fluctuation mode of $\theta_H$.

Quark-lepton multiplets and additional dark fermion multiplets are introduced in the bulk.
On the UV brane Majorana singlet fermions and brane scalar are introduced.
The matter field content of the model is summarized in Table \ref{Table:GHUmatter}.
The orbifold BCs for fermion fields in the bulk are
\begin{align}
\Psi_{({\bf 3,4})}^{q , \alpha} (x, y_j - y)  &= 
- P_{\bf 4}^{SO(5)} \gamma^5 \Psi_{({\bf 3,4})}^{q, \alpha} (x, y_j + y) ~, \cr
\Psi_{({\bf 1,4})}^{\ell , \alpha} (x, y_j - y) &= 
- P_{\bf 4}^{SO(5)} \gamma^5 \Psi_{({\bf 1,4})}^{\ell , \alpha} (x, y_j + y) ~, \cr
\Psi_{({\bf 3,1})^\pm }^{q, \alpha}  (x, y_j - y)  &=
\mp \gamma^5 \Psi_{({\bf 3,1})^\pm}^{q, \alpha}  (x, y_j + y) ~, \cr
&\alpha = 1 , 2 , 3 , 
\label{quarkleptonBC1}
\end{align}
for quark and lepton multiplets  and 
\begin{align}
&\Psi^{F_q, \alpha}_{({\bf 3},{\bf 4})}  (x, y_j - y) = 
(-1)^j  P_{\bf 4}^{SO(5)}  \gamma^5 \Psi^{F_q, \alpha}_{({\bf 3},{\bf 4})} (x, y_j + y) ~, \cr
&\Psi^{F_\ell, \alpha}_{({\bf 1},{\bf 4})}  (x, y_j - y) = 
(-1)^j  P_{\bf 4}^{SO(5)}  \gamma^5 \Psi^{F_\ell, \alpha}_{({\bf 1},{\bf 4})} (x, y_j + y) ~, \cr
&\Psi^{V, \beta}_{({\bf 1},{\bf 5})^\pm}  (x, y_j - y) =
\mp P_{\bf 5}^{SO(5)} \gamma^5 \Psi^{V, \beta}_{({\bf 1},{\bf 5})^\pm} (x, y_j + y) ~, \cr
&\qquad
 \alpha = 1, \cdots, n_F , ~~~   \beta = 1, \cdots, n_V ,
\label{DFBC1}
\end{align}
for dark fermion multiplets.

The bulk part of the action for the quark multiplets is given by
\begin{align}
I_{\rm bulk}^{\rm quark} &=  \int d^5x\sqrt{- G} \,
 \bigg\{ \sum_{\alpha = 1}^3 \Big( \overline{\Psi}{}_{\bf (3,4)}^\alpha   {\cal D} (c_{q_\alpha}) \Psi_{\bf (3,4)}^\alpha
+ \overline{\Psi}{}_{\bf (3,1)}^{+ \alpha}   {\cal D} (c_{D_\alpha} ) \Psi_{\bf (3,1)}^{+ \alpha} \cr
\noalign{\kern 5pt}
&\hskip -0.5cm
+ \overline{\Psi}{}_{\bf (3,1)}^{- \alpha}   {\cal D} (c_{D_\alpha} ) \Psi_{\bf (3,1)}^{- \alpha} \Big)
-  \sum_{\alpha, \beta=1}^3  \Big( m_{\alpha \beta}  \overline{\Psi}{}_{\bf (3,1)}^{+ \alpha} \Psi_{\bf (3,1)}^{- \beta} 
+ m_{\alpha \beta}^*  \overline{\Psi}{}_{\bf (3,1)}^{- \beta} \Psi_{\bf (3,1)}^{+ \alpha} \Big) \bigg\} ,
\label{fermionAction1}
\end{align} 
where $G = \det G_{MN}$, $\overline{\Psi} = i \Psi^\dagger \gamma^0$ and the covariant derivative ${\cal D} (c)$ is given by 
\begin{align}
&{\cal D} (c)   = \Gamma^a {e_a}^M \Big( D_M 
+ \frac{1}{8} \omega_{bcM} [\Gamma^b, \Gamma^c]  \Big) - c \,  \sigma' (y) ~,  \cr
\noalign{\kern 5pt}
&D_M =  \dd_M  -i g_A A_M -i g_B Q_X B_M - ig_C A_M^{SU(3)_C} ~. 
\label{covariantD}
\end{align}
Here $\omega_{bc} = \omega_{bcM} d x^M$ is the spin-connection 1-form.  $Q_X$ and $c$ are
$U(1)_X$ charge and  bulk mass parameter of the $\Psi$ field.
The bulk part of the action for the lepton multiplets and the action for Majorana fermions $\hat \chi^\alpha$ are 
\begin{align}
I^{\rm lepton}_{\rm bulk} &= \int d^5 x \, \sqrt{-  G} \, 
\sum_{\alpha=1}^3   \overline{\Psi}_{({\bf 1,4})}^{\alpha} \cD (c_{\ell_\alpha}) \Psi_{({\bf 1,4})}^{\alpha} ~, \cr
\noalign{\kern 5pt}
I_{\rm brane}^\chi &=
\int d^5x \sqrt{- G} \, \delta(y)  \sum_{\alpha, \beta =1}^3 \frac{1}{2}  \Big\{  
\delta_{\alpha \beta} \overline{\hat\chi}{}^\alpha \gamma^\mu\partial_\mu \hat \chi^\beta
 + M_{\alpha \beta}  \overline{\hat \chi}{}^\alpha \hat \chi^{\beta} \Big\} ~.
\label{LeptonAction1}
\end{align}


\begin{table}[tbh]
\renewcommand{\arraystretch}{1.4}
\begin{center}
\caption{The matter fields in the GUT-inspired  GHU model.
The $(SU(3)_C, SO(5))_{U(1)_X}$ content of each field is shown in the last column.}
{\vskip 10pt}
{\begin{tabular}{ccc}
\hline 
in the bulk &quark $\Psi^q$
&$({\bf 3}, {\bf 4})_{\frac{1}{6}} ~ ({\bf 3}, {\bf 1})_{-\frac{1}{3}}^+ 
    ~ ({\bf 3}, {\bf 1})_{-\frac{1}{3}}^-$\\
$(1 \le z  \le z_L)$ &lepton $\Psi^\ell$
&$\strut ({\bf 1}, {\bf 4})_{-\frac{1}{2}}$ \\
&dark fermion $\Psi^{F_q}, \Psi^{F_\ell}$ & $({\bf 3}, {\bf 4})_{\frac{1}{6}} \, ,  \, ({\bf 1}, {\bf 4})_{-\frac{1}{2}}$  \\
&dark fermion $\Psi^V$ & $({\bf 1}, {\bf 5})_{0}^+ ~ ({\bf 1}, {\bf 5})_{0}^-$  \\
\hline 
on the UV brane
&Majorana fermion $\hat \chi$ &$({\bf 1}, {\bf 1})_{0} $ \\
(at $z=1$) &brane scalar $\hat \Phi$ &$({\bf 1}, {\bf 4})_{\frac{1}{2}} $ \\
\hline 
\end{tabular}
}
\label{Table:GHUmatter}
\end{center}
\end{table}


The brane interactions of $\hat \Phi$ with quark and lepton multiplets are given by 
\begin{align}
I_{\rm brane}^{\rm int} &=  \int d^5x\sqrt{-\det G} \, \delta(y) \, \Big\{ {\cal L}_{\rm brane}^q + {\cal L}_{\rm brane}^\ell \Big\} ~, \cr
\noalign{\kern 5pt}
{\cal L}_{\rm brane}^q &=
- \Big\{ \sum_{\alpha=1}^3 \kappa^q_{\alpha} \,
\overline{\Psi}{}_{({\bf 3,4})}^{\alpha}  \hat \Phi_{({\bf 1,4})}
 \Psi_{({\bf 3,1})}^{+\alpha}  + {\rm H.c.} \Big\} ~,  \cr
 \noalign{\kern 5pt}
{\cal L}_{\rm brane}^\ell &=
 - \Big\{  \sum_{\alpha=1}^3  \kappa^\ell_{\alpha} \,
\overline{\hat \chi}{}^{\alpha}  \check \Phi_{({\bf 1,4})}^\dagger
 \Psi_{({\bf 1,4})}^{\alpha}  + {\rm H.c.} \Big\}  ~.
\label{BraneInt1}
\end{align}
Here
\begin{align}
&\hat \Phi _{\bf (1,4)} = \begin{pmatrix} \Phi_{\bf [2,1]} \cr \Phi_{\bf [1,2]}  \end{pmatrix}, ~~~
\check \Phi _{\bf (1,4)} = \begin{pmatrix} i \sigma^2  \Phi_{\bf [2,1]}^* \cr - i \sigma^2 \Phi_{\bf [1,2]}^*  \end{pmatrix} .
\label{BraneScalar1}
\end{align}
Nonvanishing $\la  \Phi_{\bf [1,2]}\ra = (0, w)^t$ spontaneously breaks $SU(2)_R \times U(1)_X$ to $U(1)_Y$.

Parity assignment and $G_{22}=SU(2)_L\times SU(2)_R$ content of quark-lepton multiplets  
are tabulated in Table \ref{Table:quarklepton}.
A consistent models is obtained for small $\theta_H$.  
As typical values we take  $\theta_H = 0.1$ and KK mass scale $m_\KK = \pi k /(z_L -1) = 13\,$TeV.
With mixing in the mass matrix of $\Psi_{({\bf 3,1})^\pm}^{q, \alpha}$ fields in the quark sector and
 mixing in Majorana mass terms for $\hat \chi^\alpha$ fields, both of the  CKM matrix and
PMNS matrix are obtained \cite{CKM2025, neutrino2026}.
The bulk mass parameters of $\Psi_{({\bf 3,4})}^{q, \alpha}$ and $\Psi_{({\bf 1,4})}^{\ell, \alpha}$
are $(c_{u}, c_{c}, c_{t}) = (-0.85912,  -0.71913,  - 0.27455)$ and
$ (c_e, c_\mu, c_\tau) = (-1.00684, -0.79302, -0.67539)$
for $\theta_H=0.1$ and $m_\KK=13\,$TeV.

\begin{table}[tbh]
\renewcommand{\arraystretch}{1.2}
\begin{center}
\caption{Parity assignment  of quark-lepton multiplets.
In the fourth column ``Left-handed'' parity assignment of left-handed components of $\Psi$ is shown.
Parity of right-handed components is opposite to that of left-handed components.
In the second column $\big( SU(3)_C, SO(5) \big)_{U(1)_X}$ content is shown.
In the third column $G_{22} =SU(2)_L\times SU(2)_R$ content is shown.
}
\vskip 10pt
\begin{tabular}{ccccc}
\hline \hline 
Field & ${\cal G}$ & $G_{22}$ &Left-handed  &Name\\
\hline
$\Psi_{({\bf 3,4})}^{q, \alpha}$ &$({\bf 3,4})_{\frac{1}{6}}$ &$\, [{\bf 2} , {\bf  1}] \,$
&$(+,+)$  &$u ~~~ c~~~ t$\\
&&&& $d ~~~ s ~~~ b$\\
&&$[{\bf 1} , {\bf  2}]$ 
&$(-,-)$ &$u' ~~~ c'~~~ t'$\\
&&&& $d' ~~~ s'~~~ b'$\\
$\Psi_{({\bf 3,1})^\pm}^{q, \alpha}$ &$({\bf 3,1})_{-\frac{1}{3}}$ 
&$[{\bf 1} , {\bf  1}]$
&$(\pm ,\pm )$  &$D^{\pm}_d ~ D^{\pm}_s ~ D^{\pm}_b$\\
\noalign{\kern 5pt}
\hline
\noalign{\kern 5pt}
$\Psi_{({\bf 1,4})}^{\ell, \alpha}$ &$({\bf 1,4})_{- \frac{1}{2}}$ &$\, [{\bf 2} , {\bf  1}] \,$
&$(+,+)$ &$\nu_e \,~  \nu_\mu ~\,  \nu_\tau$\\
&&&& $e ~~~ \mu ~~~ \tau$\\
&&$[{\bf 1} , {\bf  2}]$ 
&$(-,-)$  &$\nu_e' \,~  ~ \nu_\mu' ~\,  \nu_\tau'$\\
&&&& $e' ~~~ \mu' ~~ \tau'$\\
\hline
$\hat \chi^\alpha$ &$({\bf 1,1})_0$ 
&$[{\bf 1} , {\bf  1}]$
&$\cdots$  &$\eta_e ~\, \eta_\mu ~\, \eta_\tau$\\
\hline \hline
\end{tabular}
\label{Table:quarklepton}
\end{center}
\end{table}

\section{Baryon and lepton numbers}

The action is invariant under a global $U(1)_B$ transformation
\begin{align}
\Psi_{({\bf 3,4})}^{q, \alpha} (x,y) &\go e^{i \ep} \,  \Psi_{({\bf 3,4})}^{q, \alpha} (x,y) ~, \cr
\noalign{\kern 5pt}
\Psi_{({\bf 3,1})^\pm }^{q, \alpha} (x,y) &\go e^{i \ep} \,  \Psi_{({\bf 3,1})^\pm}^{q, \alpha} (x,y) ~.
\label{Baryon1}
\end{align}
Let us write the whole action as 
$I = \int d^5 x \sqrt{-G} \big\{ {\cal L}_{\rm bulk} + \delta (y) {\cal L}_{\rm brane} \big\} = \int d^5 x \, {\cal L}_0$.
The 5D Noether current $J_B^M (x,y)$ following from the $U(1)_B$ invariance (\ref{Baryon1}) is
\begin{align}
J_B^M (x,y) &= \frac{i}{\sqrt{-G}}  \sum_{\Psi^\beta = \Psi_{({\bf 3,4})}^{q, \alpha}  , \, \Psi_{({\bf 3,1})^\pm }^{q, \alpha} }
\bigg\{ \frac{\delta{\cal L}_0}{\delta \dd_M \Psi^\beta} \Psi^\beta 
- \overline{\Psi}^\beta \frac{\delta{\cal L}_0}{\delta \dd_M \overline{\Psi}^\beta} \bigg\} \cr
\noalign{\kern 5pt}
&= i {e_a}^M \sum_{\alpha=1}^3 \Big\{ \overline{\Psi}_{({\bf 3,4})}^{q, \alpha} \Gamma^a \Psi_{({\bf 3,4})}^{q, \alpha} 
+ \overline{\Psi}_{({\bf 3,1})^+ }^{q, \alpha}  \Gamma^a \Psi_{({\bf 3,1})^+ }^{q, \alpha} 
+ \overline{\Psi}_{({\bf 3,1})^- }^{q, \alpha}  \Gamma^a \Psi_{({\bf 3,1})^- }^{q, \alpha} \Big\} ~.
\label{Baryon2}
\end{align}
At the classical level the current is conserved, 
but an anomaly term arises at the quantum level.
\begin{align}
{J_B^M}_{; M} &= \frac{1}{\sqrt{-G}} \dd_M \Big( \sqrt{-G}  J_B^M \Big) = {\cal F}_B^{\rm anom} ~.
\label{BaryonAnom1}
\end{align}

Baryon number $B$  is given by
\begin{align}
B &= \int d^3 x \int_0^L dy \, \sqrt{-G} \, J_B^0 (x,y) \cr
\noalign{\kern 5pt}
&=  \int d^3 x \int_1^{z_L} dz \, \frac{1}{k} \sum_{\alpha=1}^3  
\Big\{ \check{\Psi}_{({\bf 3,4})}^{q, \alpha} {}^\dagger  \check{\Psi}_{({\bf 3,4})}^{q, \alpha} 
+ \check{\Psi}_{({\bf 3,1})^+}^{q, \alpha} {}^\dagger  \check{\Psi}_{({\bf 3,1})^+}^{q, \alpha} 
+ \check{\Psi}_{({\bf 3,1})^-}^{q, \alpha} {}^\dagger  \check{\Psi}_{({\bf 3,1})^-}^{q, \alpha} \Big\} 
\label{Baryon3}
\end{align}
where $\check \Psi = z^{-2}\Psi$.
Inserting the KK expansions of $\Psi$'s given in the next section, one finds for the first generation ($\alpha=1$)
\begin{align}
B^{\alpha=1} &= \int d^3 x \bigg\{ \sum_{n=0}^\infty \big( u^{(n) \dagger }u^{(n)} + d^{(n)\dagger} d^{(n)} \big)
+ \sum_{n=1}^\infty D_d^{(n) \dagger}  D_d^{(n)} \bigg\}.
\label{Baryon4}
\end{align}
We note that 
\begin{align}
\frac{d B}{dx_0} &=  \frac{1}{2} \int d^3 x \int_{-L}^L dy \,  \Big\{ - \dd_y \Big(  \sqrt{-G} \, J_B^y \Big) 
+ \sqrt{-G} \,  {\cal F}_B^{\rm anom}  \Big\} \cr
\noalign{\kern 5pt}
&=  \frac{1}{2} \int d^3 x \int_{-L}^{L}  dy \,  \sqrt{-G} \,  {\cal F}_B^{\rm anom}   ~.
\end{align}
Here we made use of a fact that $\dd_y \big(  \sqrt{-G} \, J_B^y \big)$ is parity even and periodic in $y$ with a period $2L$.

The baryon number current in four dimensions, $j_B^\mu (x)$,  is given by
\begin{align}
j_B^\mu (x) &= \int_0^L dy \, \sqrt{-G} \, J_B^\mu (x,y) \cr
\noalign{\kern 5pt}
&= \frac{i}{k} \int_1^{z_L} dz    \sum_{\alpha=1}^3  
\Big\{ \bar{\check{\Psi}}_{({\bf 3,4})}^{q, \alpha} \gamma^\mu  \check{\Psi}_{({\bf 3,4})}^{q, \alpha} 
+ \bar{\check{\Psi}}_{({\bf 3,1})^+}^{q, \alpha} \gamma^\mu \check{\Psi}_{({\bf 3,1})^+}^{q, \alpha} 
+ \bar{\check{\Psi}}_{({\bf 3,1})^-}^{q, \alpha} \gamma^\mu \check{\Psi}_{({\bf 3,1})^-}^{q, \alpha} \Big\} 
\label{4Dbaryon1}
\end{align}
which satisfies 
\begin{align}
\dd_\mu j_B^\mu (x) &= \int_0^L dy \,  \sqrt{-G} \,  {\cal F}_B^{\rm anom}   ~.
\label{4Dbaryon2}
\end{align}
For the first generation
\begin{align}
j_B^\mu (x)^{\alpha=1} &= \sum_{n=0}^\infty \Big( \bar{u}^{(n) } i \gamma^\mu u^{(n)} 
+ \bar{d}^{(n)}  i \gamma^\mu d^{(n)} \Big)
+ \sum_{n=1}^\infty \bar{D}_d^{(n)}   i \gamma^\mu D_d^{(n)} ~.
\label{4Dbaryon3}
\end{align}
In the following sections we determine ${\cal F}_B^{\rm anom}$ by applying (\ref{4Dbaryon2}) and  (\ref{4Dbaryon3}).

In the lepton sector we consider a global $U(1)_L$ transformation
\begin{align}
\Psi_{({\bf 1,4})}^{\ell, \alpha} (x,y) &\go e^{i \ep} \,  \Psi_{({\bf 1,4})}^{\ell, \alpha} (x,y)  ~.
\label{Lepton1}
\end{align}
The brane interaction ${\cal L}_{\rm brane}^\ell$ 
in (\ref{BraneInt1}), however, is not invariant under $U(1)_L$ transformations.
The presence of Majorana brane fermions $\hat \chi^\alpha$ necessarily breaks $U(1)_L$.
5D lepton number current $J_L^M (x,y)$  is given by
\begin{align}
J_L^M (x,y) &= \frac{i}{\sqrt{-G}}  \sum_{\alpha=1}^3
\bigg\{ \frac{\delta{\cal L}_0}{\delta \dd_M \Psi_{({\bf 1,4})}^{\ell, \alpha} } \Psi_{({\bf 1,4})}^{\ell, \alpha} 
- \overline{\Psi}_{({\bf 1,4})}^{\ell, \alpha} 
\frac{\delta{\cal L}_0}{\delta \dd_M \overline{\Psi}_{({\bf 1,4})}^{\ell, \alpha} } \bigg\} \cr
\noalign{\kern 5pt}
&= i {e_a}^M \sum_{\alpha=1}^3  \overline{\Psi}_{({\bf 1,4})}^{\ell, \alpha}
 \Gamma^a \Psi_{({\bf 1,4})}^{\ell, \alpha} ~.
 \label{Lepton2}
\end{align}
This current satisfies, unlike Eq.\ (\ref{BaryonAnom1}), 
\begin{align}
{J_L^M}_{; M} &= \delta (y) {\cal K}_{\rm break} +  {\cal F}_L^{\rm anom} ~,  \cr
\noalign{\kern 5pt}
 {\cal K}_{\rm break} &= -i \sum_{\alpha=1}^3  \Big\{  \kappa^\ell_{\alpha} \,
\overline{\hat \chi}{}^{\alpha}  \check \Phi_{({\bf 1,4})}^\dagger  \Psi_{({\bf 1,4})}^{\alpha}  
- \kappa^\ell_{\alpha} {}^* \,  \overline{\Psi} {}_{({\bf 1,4})}^{\alpha}  \check \Phi_{({\bf 1,4})} \hat\chi^\alpha \Big\} .
\label{LeptonAnom1}
\end{align}
The lepton number current in four dimensions is
\begin{align}
j_L^\mu (x) &= \int_0^L dy \, \sqrt{-G} \, J_L^\mu (x,y) \cr
\noalign{\kern 5pt}
&= \frac{i}{k} \int_1^{z_L} dz    \sum_{\alpha=1}^3  
\bar{\check{\Psi}}_{({\bf 1,4})}^{\ell, \alpha} \gamma^\mu  \check{\Psi}_{({\bf 1,4})}^{\ell, \alpha} 
\label{4Dlepton1}
\end{align}
which satisfies 
\begin{align}
\dd_\mu j_L^\mu (x) &=\frac{1}{2} \,  {\cal K}_{\rm break} |_{y=0} 
+ \int_{0}^L dy \,  \sqrt{-G} \,  {\cal F}_L^{\rm anom}   ~.
\label{4Dlepton2}
\end{align}

\section{Anomalies}

One unambiguous approach for determining ${\cal F}_B^{\rm anom}$ in Eqs.\ (\ref{BaryonAnom1}) 
and (\ref{4Dbaryon2})  is to evaluate such matrix elements  as
$\la W^{(n)} (p) W^{\dagger (m)} (k) | j_B^\mu (x) | 0 \ra$ in perturbation theory
where $W^{(n)}$ ($n=0,1,2, \cdots$) are KK excited states of the $W$ boson tower.
KK expansions of relevant gauge bosons in $SO(5) \times U(1)_X$  
are the following.  Ten $SO(5)$ generators are decomposed to $\{ T^{a_L}, a_L=1,2,3 \}$ in $SU(2)_L$, 
$\{ T^{a_R}, a_R=1,2,3 \}$ in $SU(2)_R$, and the rest $\{ T^{\hat p}, p=1 \sim 4 \}$.
The corresponding gauge fields are denoted as $A_M^{a_L}$, $A_M^{a_R}$ and $A_M^{\hat p}$.
KK towers of $W$ and $W_R$ bosons are contained in
\begin{align}
&\frac{1}{\sqrt{2 k}} \begin{pmatrix}  A_\mu^{1_L} + i A_\mu^{2_L} \cr \mysnoalign 
A_\mu^{1_R} + i A_\mu^{2_R} \cr  \mysnoalign
A_\mu^{\hat 1} + i A_\mu^{\hat 2} \end{pmatrix} 
=  \sum_{n=0}^\infty W_\mu^{ (n)} (x) \begin{pmatrix}  h^L_{W^{(n)}} (y) \cr \mysnoalign
 h^R_{W^{(n)}} (y) \cr  \mysnoalign \hat h_{W^{(n)}} (y) \end{pmatrix} 
 + \sum_{n=1}^\infty W_{R \mu}^{ (n)} (x) \begin{pmatrix}  h^L_{W_R^{(n)}} (y) \cr 
 h^R_{W_R^{(n)}} (y) \cr  \hat h_{W_R^{(n)}} (y) \end{pmatrix} .
\label{Wtower1}
\end{align}
KK towers of $Z$, $Z_R$ and $\gamma$ bosons are contained in
\begin{align}
\frac{1}{\sqrt{k}} \begin{pmatrix}  A_\mu^{3_L}  \cr \mytinynoalign   A_\mu^{3_R}\cr 
\mytinynoalign   A_\mu^{\hat 3}  \cr B_\mu  \end{pmatrix} 
&= \sum_{n=0}^\infty Z_\mu^{ (n)} (x) \begin{pmatrix}  h^L_{Z^{(n)}} (y) \cr \mysnoalign
 h^R_{Z^{(n)}} (y) \cr  \mysnoalign    \hat h_{Z^{(n)}} (y) \cr  \mysnoalign h^B_{Z^{(n)}} (y) \end{pmatrix} 
+ \sum_{n=1}^\infty Z_{R\mu}^{ (n)} (x) \begin{pmatrix}  h^L_{Z_R^{(n)}} (y) \cr 
 h^R_{Z_R^{(n)}} (y) \cr    \hat h_{Z_R^{(n)}} (y) \cr h^B_{Z_R^{(n)}} (y) \end{pmatrix}  \cr
\noalign{\kern 5pt}
&
+\sum_{n=0}^\infty \gamma_\mu^{ (n)} (x) \begin{pmatrix} \sin \theta_W^0 \cr   \sin \theta_W^0 \cr     0 \cr 
\sqrt{1 - 2 \sin^2 \theta_W^0} \end{pmatrix} h_{\gamma^{(n)}} (y) , 
\label{Ztower1}
\end{align}
where $\sin \theta_W^0 = g_B/\sqrt{g_A^2 + 2 g_B^2}$ corresponds to $\sin \theta_W^\SM$ in the SM.
$A_\mu^{\hat 4}$ turns out irrelevant to ${\cal F}_B^{\rm anom}$. 
$W^{(0)}$, $Z^{(0)}$ and $\gamma^{(0)}$ are $W$ boson, $Z$ boson and photon, respectively.
Details of mode functions $h (y)$ are given in ref.\ \cite{YHbook} and in Appendix B.
For later use we write
\begin{align}
\begin{pmatrix}  h^L_{Z^{(n)}} (y) \cr \mysnoalign
 h^R_{Z^{(n)}} (y) \cr  \mysnoalign    \hat h_{Z^{(n)}} (y) \cr  \mysnoalign h^B_{Z^{(n)}} (y) \end{pmatrix} 
 &= \begin{pmatrix}  h^{L, su2}_{Z^{(n)}} (y) \cr \mysnoalign
 h^{R, su2}_{Z^{(n)}} (y) \cr  \mysnoalign    \hat h^{su2} _{Z^{(n)}} (y) \cr  \mysnoalign 0 \end{pmatrix} 
 - \sin \theta_W^0  \begin{pmatrix} \sin \theta_W^0 \cr   \sin \theta_W^0 \cr     0 \cr 
\sqrt{1 - 2 \sin^2 \theta_W^0} \end{pmatrix} h^{em}_{Z^{(n)}} (y) .
\label{Ztower2}
\end{align}

\begin{figure}[tbh]
\centering
\includegraphics[height=25mm]{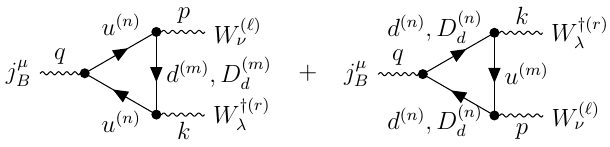}
\caption{Anomaly diagrams for the $j_B^\mu  \, W_\nu^{(\ell)} W_\lambda^{\dagger (r)} $ vertex.
}
\label{fig:WW}
\end{figure}

We consider matrix elements 
\begin{align}
&\int d^4 x \, e^{-iqx}  
 \la X^{(\ell)} (p) Y^{(r)} (k) | j_B^\mu (x) | 0\ra \cr
\noalign{\kern 5pt}
&\hskip 1.cm 
= (2\pi)^4  \delta^{(4)} (p + k -q)  \ep_\nu^*(p) \ep_\lambda^* (k)
{\cal M}^{\mu\nu\lambda}_{X^{(\ell)} Y^{(r)} }(p,k)    
\label{XY1}
\end{align}
for gauge fields $X^{(\ell)}$ and $Y^{(r)}$
where $\ep_\nu (p)$ and  $\ep_\lambda (k)$ are polarization vectors of the $X^{(\ell)}$ and $Y^{(r)} $ states.
Anomaly terms for $\dd_\mu j_B^\mu (x)$ appear as nonvanishing $i q_\mu {\cal M}^{\mu\nu\lambda}_{X^{(\ell)} Y^{(r)} }(p,k)$.
Let us first consider the matrix element for $X^{(\ell)} = W^{(\ell)}$ and $Y^{(r)} = W^{\dagger (r)}$.
One loop diagrams contributing to $i q_\mu {\cal M}^{\mu\nu\lambda}_{W^{(\ell)} W^{\dagger (r)} }(p,k) $ are depicted 
in Fig.\ \ref{fig:WW}.
In ref.\ \cite{anomaly2026} gauge anomalies in the GUT-inspired GHU model were analyzed.  One recognizes that
the diagrams in Fig.\ \ref{fig:WW} have the same structure as the  $\gamma W W^\dagger$  diagrams  in 
ref.\ \cite{anomaly2026}.  The results obtained in \cite{anomaly2026} are utilized by replacing the photon coupling
by $j_B^\mu$.

The $W u \tilde d$ couplings, where $\tilde d = d$ and $D_d$, are given by
\begin{align}
&{\cal L}^{W u \tilde d} = -i \frac{g_w}{\sqrt{2}} \sum_{\ell=0}^\infty  W_\mu^{(\ell) \dagger} 
 \sum_{n, m }^\infty  \Big\{ 
\hat g^{W u\tilde d}_{L\, \ell n m} \bar u_L^{(n)} \gamma^\mu \tilde d_L^{(m)}
+ \hat g^{W u \tilde d}_{R\, \ell n m} \bar u_R^{(n)} \gamma^\mu \tilde d_R^{(m)} \Big\} + {\rm H.c.} ~, \cr
\noalign{\kern 5pt}
&\hat g^{W u \tilde d}_{L/R\, \ell n m} = \frac{k}{2} \sqrt{kL} \int_{-L}^{L} dy \, e^{\sigma (y)} \, 
 {\cal K}_W [(h^L, h^R,  {\hat h})_{W^{(\ell)}};  ( f, g)^{u^{(n)}}_{L/R}, ( f, g)^{\tilde d^{(m)}}_{L/R} ] ~, \cr
\noalign{\kern 5pt}
&{\cal K}_W [(h^L, h^R, \hat h)_\alpha; (f, g)_1, (f, g)_2]  = h^{L*}_\alpha   f_1^*   f_2 +  h^{R*}_\alpha  g_1^*   g_2
+  \frac{i}{\sqrt{2}} \,  \hat h^*_\alpha  ( f_1^*  g_2 -  g_1^*  f_2 )  ~.
\label{Wud1}
\end{align}
It is understood that $\hat g^{W u  D_d}_{L/R\, \ell n 0} = 0$.
The KK expansions of $u$, $d$, $D_d$ towers are given by
\begin{align}
& \begin{pmatrix} {\check u} \cr {\check u}' \end{pmatrix} = 
\sqrt{k} \sum_{n=0}^\infty \bigg\{ u^{(n)}_L (x) \begin{pmatrix} f^{u^{(n)}}_L (y) \cr  g^{u^{(n)}}_L (y) \end{pmatrix}
+ u^{(n)}_R (x) \begin{pmatrix}  f^{u^{(n)}}_R (y) \cr  g^{u^{(n)}}_R (y) \end{pmatrix} \bigg\} , 
\label{wave-up1}
\end{align}
and 
\begin{align}
 \begin{pmatrix}\check  d \cr \check d'  \cr \check D_d^+ \cr \check D_d^- \end{pmatrix} 
&=  \sqrt{k} \sum_{n=0}^\infty \Bigg\{ d^{(n)}_L (x) 
\begin{pmatrix} f^{d^{(n)}}_L (y) \cr g^{d^{(n)}}_L (y) \cr h^{d^{(n)}}_L (y) \cr k^{d^{(n)}}_L (y)\end{pmatrix}
+ d^{(n)}_R (x) 
\begin{pmatrix} f^{d^{(n)}}_R (y) \cr g^{d^{(n)}}_R (y)  \cr h^{d^{(n)}}_R (y) \cr k^{d^{(n)}}_R (y) \end{pmatrix} \Bigg\} \cr
\noalign{\kern 5pt}
&\hskip -.3cm 
+  \sqrt{k} \sum_{n=1}^\infty \Bigg\{ D^{(n)}_{d L} (x) 
\begin{pmatrix} f^{D_d^{(n)}}_L (y) \cr g^{D_d^{(n)}}_L (y) \cr h^{D_d^{(n)}}_L (y) \cr k^{D_d^{(n)}}_L (y)\end{pmatrix}
+ D^{(n)}_{dR} (x) 
\begin{pmatrix} f^{D_d^{(n)}}_R (y) \cr g^{D_d^{(n)}}_R (y)  \cr h^{D_d^{(n)}}_R (y) \cr k^{D_d^{(n)}}_R (y) \end{pmatrix} \Bigg\} .
\label{wave-down1}
\end{align}
Details of mode functions  are given in refs.\ \cite{YHbook} and  \cite{anomaly2026}. 

By standard manipulations of diagrams in Fig.\ \ref{fig:WW} one finds, for $u d$ loops,  that 
\begin{align}
i q_\mu  {\cal M}^{\mu\nu\lambda}_{W^{(\ell)} W^{\dagger (r)} }(p,k)^{ud}
&= - \frac{g_w^2}{8 \pi^2} \, \ep^{\nu \lambda \alpha \beta} \, p_\alpha k_\beta \, F_{W^{(\ell)} W^{\dagger (r)} }^{ud} ~, \cr
\noalign{\kern 5pt}
F_{W^{(\ell)} W^{\dagger (r)} }^{ud} &=  \sum_{\tilde d = d, D_d}
\Big( \Tr \hat g^{Wu \tilde d}_{L \, r}  \hat g^{W^\dagger \tilde d u}_{L \, \ell}  
- \Tr \hat g^{Wu \tilde d}_{R \, r}  \hat g^{W^\dagger \tilde d u}_{R \, \ell}  \Big)  ~,  \cr
\noalign{\kern 5pt}
\big( \hat g^{Wu \tilde d}_{L/R \, r}  \big)_{nm} &= \hat g^{Wu \tilde d}_{L/R \, r nm}  ~, ~~
\big( \hat g^{W^\dagger  \tilde d u}_{L/R \, \ell} \big)_{mn}  = \big( \hat g^{Wu \tilde d}_{L/R \, \ell nm} \big)^* ~.
\label{anomWW1}
\end{align}
The factor $F_{W^{(\ell)} W^{\dagger (r)}}^{ud}$ is evaluated in two manners.
One way is to evaluate the couplings $\hat g^{Wu \tilde d}_{L/R \, \ell nm} $ up to the  KK level  $n_\max$,  
namely for $n, m \le n_\max$, 
and do the matrix trace $\Tr$. 
In the $n_\max \go \infty$ limit  $F_{W^{(\ell)} W^{\dagger (r)}}^{ud}$ is found.
A second way is to first write $F_{W^{(\ell)} W^{\dagger (r)}}^{ud}$, with the aid of Eq.\ (\ref{Wud1}),  as
\begin{align}
F_{W^{(\ell)} W^{\dagger (r)} }^{ud} &=  \sum_{\tilde d = d, D_d}
 \Big( \frac{k}{2} \sqrt{kL} \Big)^2 \int_{-a}^{2L -a} dy_1  
\int_{-a}^{2L -a} dy_2  \, e^{\sigma (y_1) + \sigma (y_2)}  \sum_{n,m=0}^\infty  \cr
\noalign{\kern 5pt}
&
\times \Big\{ {\cal K}_{W^{(r)}}^{u^{(n)}_{L}, \tilde d^{(m)}_{L} } (y_1)  {\cal K}_{W^{(\ell)}}^{u^{(n)}_{L}, \tilde d^{(m)}_{L} } (y_2)^*
- {\cal K}_{W^{(r)}}^{u^{(n)}_{R}, \tilde d^{(m)}_{R} } (y_1)  {\cal K}_{W^{(\ell)}}^{u^{(n)}_{R}, \tilde d^{(m)}_{R} } (y_2)^* \Big\} , \cr
\noalign{\kern 5pt}
{\cal K}_{W^{(r)}}^{u^{(n)}_{L/R}, \tilde d^{(m)}_{L/R} } (y) &=
 {\cal K}_W [(h^L, h^R,  {\hat h})_{W^{(r)}};  ( f, g)^{u^{(n)}}_{L/R}, ( f, g)^{\tilde d^{(m)}}_{L/R} ] (y) ~,
 \label{anomWW2}
\end{align}
and do the KK sum $\sum_{n,m=0}^\infty$ first and then perform the integrations over $y_1$ and $y_2$.

In ref.\ \cite{anomaly2026} the $F$ factors were evaluated for $\ell=r=0$ in both methods.
It was shown that $F_{W^{(0)} W^{\dagger (0)} }^{ud} = F_{W^{(0)} W^{\dagger (0)} }^{cs} = F_{W^{(0)} W^{\dagger (0)} }^{tb} $.
Further it was shown that  $F_{W^{(0)} W^{\dagger (0)} }^{ud} $ is expressed in terms of the values of the $W^{(0)}$ 
wave functions at the UV and IR branes.
These properties remain valid for general $\ell$ and $r$.  
Indeed one can confirm numerically in the first method that
\begin{align}
&F_{W^{(\ell)} W^{\dagger (r)} }^{ud} = F_{W^{(\ell)} W^{\dagger (r)} }^{cs} = F_{W^{(\ell)} W^{\dagger (r)} }^{tb} 
\equiv  F_{W^{(\ell)} W^{\dagger (r)} } ~.
\label{anomWW3}
\end{align}
Although fermion wave functions, $(f,g)^{u^{(n)}}_{L/R}$ etc.  and gauge couplings 
$\hat g^{Wu \tilde d}_{L/R \, r nm}$ etc. depend on fermion species, the $F$ factors are universal.
Holographic formulas for the $F$ factors are derived in the second method.  The argument for $F_{W^{(0)} W^{\dagger (0)}} $
given in ref.\ \cite{anomaly2026} is easily generalized for $F_{W^{(\ell)} W^{\dagger (r)}}$.
Taking advantage of the universality relations in Eq.\ (\ref{anomWW3}), $F_{W^{(\ell)} W^{\dagger (r)}}$ is evaluated in the case 
with the bulk mass parameter $c_u$ of fermions vanishing and brane interactions for $d$-$D_d$ being absent
so that ${\cal K}_{W^{(r)}}^{u^{(n)}_{L/R}, {D_d}^{(m)}_{L/R} } = 0$.
In taking the KK sums $\sum_n$ and $\sum_m$ in Eq.\ (\ref{anomWW2}), one encounters expressions such as
\begin{align}
&\begin{pmatrix} A_{L/R}^{uu}  \cr B_{L/R}^{uu} \cr C_{L/R}^{uu} \cr D_{L/R}^{uu} \end{pmatrix}  (y_j, y_k)
= \sum_{n=0}^\infty \begin{pmatrix} f_{L/R}^{u^{(n)}} (y_j) f_{L/R}^{u^{(n)}} (y_k)^* \cr
g_{L/R}^{u^{(n)}} (y_j) g_{L/R}^{u^{(n)}} (y_k)^* \cr f_{L/R}^{u^{(n)}} (y_j) g_{L/R}^{u^{(n)}} (y_k)^* \cr 
g_{L/R}^{u^{(n)}} (y_j) f_{L/R}^{u^{(n)}} (y_k)^* \end{pmatrix} .
\label{completeness1}
\end{align}
These sums are found to be, for $c=0$, \cite{anomaly2026, AnomalyFlow2}
\begin{align}
&A_L(y_1, y_2) = B_R(y_1, y_2) 
= \frac{e^{-\sigma (y_1)}}{k} \big\{ \delta_{2L} (y_1 - y_2) + \delta_{2L} (y_1 + y_2) \big\} , \cr
\noalign{\kern 5pt}
&A_R(y_1, y_2) = B_L (y_1, y_2) 
= \frac{e^{-\sigma (y_1)}}{k} \big\{ \delta_{2L} (y_1 - y_2) - \delta_{2L} (y_1 + y_2) \big\} , \cr
\noalign{\kern 5pt}
&C_{L/R} (y_1, y_2) = D_{L/R} (y_1, y_2) = 0 ~, 
\label{ABsumrelation1}
\end{align}
where $\delta_L (x) = \sum_{n=-\infty}^\infty \delta (x - nL)$.
Inserting (\ref{ABsumrelation1}) into (\ref{anomWW2}), one finds 
\begin{align}
F_{W^{(\ell)} W^{\dagger (r)} } &=  \Big( \frac{k}{2} \sqrt{kL} \Big)^2 \int_{-a}^{2L -a} dy_1  
\int_{-a}^{2L -a} dy_2  \, e^{\sigma (y_1) + \sigma (y_2)} \cr
\noalign{\kern 5pt}
&
\times
\Big(  h^{L \, *}_1 h^{L}_2 \big\{ A_L^{2,1}  A_L^{1,2}   - A_R^{2,1}  A_R^{1,2} \big\}  
+  h^{R \, *}_1 h^{R}_2 \big\{ B_L^{2,1}  B_L^{1,2}   - B_R^{2,1}  B_R^{1,2} \big\}   \cr
\noalign{\kern 5pt}
&\hskip 1.cm 
+ \frac{1}{2} \hat h_1^*  \hat h_2
\big\{ A_L^{2,1}  B_L^{1,2}  + B_L^{2,1}  A_L^{1,2}   - A_R^{2,1}  B_R^{1,2} - B_R^{2,1}  A_R^{1,2}  \big\} \Big)  \cr
\noalign{\kern 5pt}
&= \frac{kL}{2} \Big\{  h^{L}_{W^{(r)}} (0)^* \, h^{L}_{W^{(\ell)}} (0)  - h^{R}_{W^{(r)}} (0)^* \,  h^{R}_{W^{(\ell)}} (0) \cr
\noalign{\kern 5pt}
&\hskip 1.3cm 
+ h^{L}_{W^{(r)}} (L)^* \, h^{L}_{W^{(\ell)}} (L)  - h^{R}_{W^{(r)}} (L)^* \,  h^{R}_{W^{(\ell)}} (L) \Big\}  ~, 
\label{Fidentity5}
\end{align}
where $h^{L/R}_1 =  h^{L/R}_{W^{(r)}}  (y_1)$,  $h^{L/R}_2 =  h^{L/R}_{W^{(\ell)}}  (y_2)$, 
$A^{1,2}_{L/R} = A_{L/R} (y_1, y_2)$ etc.   The $\hat h_1^*  \hat h_2$ term vanishes.
It is seen that $F_{W^{(\ell)} W^{\dagger (r)} }$ is expressed in terms of the values of the wave functions of
$W^{(\ell)}$ and $W^{(r)}$ at $y=0$ and $L$.

Similarly the $W_R u \tilde d$ couplings are given by
\begin{align}
&{\cal L}^{W_R u \tilde d} = -i \frac{g_w}{\sqrt{2}} \sum_{\ell=1}^\infty  W_{R \mu}^{(\ell) \dagger} 
 \sum_{n, m }^\infty  \Big\{ 
\hat g^{W_R u\tilde d}_{L\, \ell n m} \bar u_L^{(n)} \gamma^\mu \tilde d_L^{(m)}
+ \hat g^{W_R u \tilde d}_{R\, \ell n m} \bar u_R^{(n)} \gamma^\mu \tilde d_R^{(m)} \Big\} + {\rm H.c.} ~, \cr
\noalign{\kern 5pt}
&\hat g^{W_R u \tilde d}_{L/R\, \ell n m} = \frac{k}{2} \sqrt{kL} \int_{-L}^{L} dy \, e^{\sigma (y)} \, 
 {\cal K}_W [(h^L, h^R, 0)_{W_R^{(\ell)}};  ( f, g)^{u^{(n)}}_{L/R}, ( f, g)^{\tilde d^{(m)}}_{L/R} ] ~.
\label{WRud1}
\end{align}
One loop diagrams contributing to  $i q_\mu {\cal M}^{\mu\nu\lambda}_{W_R^{(\ell)} W_R^{\dagger (r)} }(p,k)$ 
are diagrams obtained from those in Fig.\ \ref{fig:WW} by replacing $W_\nu^{(\ell)}$ and $W_\lambda^{\dagger (r)}$
 by $W_{R\, \nu}^{(\ell)}$ and $W_{R\, \lambda}^{\dagger (r)}$, respectively.
Evaluation of  $i q_\mu {\cal M}^{\mu\nu\lambda}_{W_R^{(\ell)} W_R^{\dagger (r)} }(p,k)$ and
$i q_\mu {\cal M}^{\mu\nu\lambda}_{W_R^{(\ell)} W^{\dagger (r)} }(p,k)$ proceeds along the same line 
as that for $i q_\mu {\cal M}^{\mu\nu\lambda}_{W^{(\ell)} W^{\dagger (r)} }(p,k)$ given above.
\begin{align}
i q_\mu  {\cal M}^{\mu\nu\lambda}_{W_R^{(\ell)} W_R^{\dagger (r)} }(p,k)^{ud}
&= - \frac{g_w^2}{8 \pi^2} \, \ep^{\nu \lambda \alpha \beta} \, p_\alpha k_\beta \, F_{W_R^{(\ell)} W_R^{\dagger (r)} }^{ud} ~, \cr
\noalign{\kern 5pt}
i q_\mu  {\cal M}^{\mu\nu\lambda}_{W_R^{(\ell)} W^{\dagger (r)} }(p,k)^{ud}
&= - \frac{g_w^2}{8 \pi^2} \, \ep^{\nu \lambda \alpha \beta} \, p_\alpha k_\beta \, F_{W_R^{(\ell)} W^{\dagger (r)} }^{ud} ~, \cr
\noalign{\kern 5pt}
F_{W_R^{(\ell)} W_R^{\dagger (r)} }^{ud} &=  \sum_{\tilde d = d, D_d}
\Big( \Tr \hat g^{W_R u \tilde d}_{L \, r}  \hat g^{W_R ^\dagger \tilde d u}_{L \, \ell}  
- \Tr \hat g^{W_R u \tilde d}_{R \, r}  \hat g^{W_R^\dagger \tilde d u}_{R \, \ell}  \Big)  ~,  \cr
\noalign{\kern 5pt}
F_{W_R^{(\ell)} W^{\dagger (r)} }^{ud} &=  \sum_{\tilde d = d, D_d}
\Big( \Tr \hat g^{W u \tilde d}_{L \, r}  \hat g^{W_R^\dagger \tilde d u}_{L \, \ell}  
- \Tr \hat g^{W u \tilde d}_{R \, r}  \hat g^{W_R^\dagger \tilde d u}_{R \, \ell}  \Big)  ~,  \cr
\noalign{\kern 5pt}
\big( \hat g^{W_R u \tilde d}_{L/R \, r}  \big)_{nm} &= \hat g^{W_R u \tilde d}_{L/R \, r nm}  ~, ~~
\big( \hat g^{W_R^\dagger  \tilde d u}_{L/R \, \ell} \big)_{mn}  = \big( \hat g^{W_R u \tilde d}_{L/R \, \ell nm} \big)^* ~.
\label{anomWWR1}
\end{align}
As in Eq.\ (\ref{anomWW3}) universality relations
\begin{align}
&F_{W_R^{(\ell)} W_R^{\dagger (r)} }^{ud} = F_{W_R^{(\ell)} W_R^{\dagger (r)} }^{cs} =
 F_{W_R^{(\ell)} W_R^{\dagger (r)} }^{tb}  \equiv  F_{W_R^{(\ell)} W_R^{\dagger (r)} } ~, \cr
\noalign{\kern 5pt}
&F_{W_R^{(\ell)} W^{\dagger (r)} }^{ud} = F_{W_R^{(\ell)} W^{\dagger (r)} }^{cs} =
 F_{W_R^{(\ell)} W^{\dagger (r)} }^{tb}  \equiv  F_{W_R^{(\ell)} W^{\dagger (r)} } ~,
\label{anomWWR2}
\end{align}
hold.
Since $h^{L/R}_{W_R^{(\ell)}} (0) = 0$, we find
\begin{align}
F_{W_R^{(\ell)} W_R^{\dagger (r)}}  &= \frac{kL}{2} \Big\{ h^{L}_{W_R^{(r)}} (L)^* \, h^{L}_{W_R^{(\ell)}} (L)  
- h^{R}_{W_R^{(r)}} (L)^* \,  h^{R}_{W_R^{(\ell)}} (L) \Big\} ~,   \cr
\noalign{\kern 5pt}
F_{W_R^{(\ell)} W^{\dagger (r)}}  &= \frac{kL}{2} \Big\{ h^{L}_{W^{(r)}} (L)^* \, h^{L}_{W_R^{(\ell)}} (L)  
- h^{R}_{W^{(r)}} (L)^* \,  h^{R}_{W_R^{(\ell)}} (L) \Big\} ~.  
\label{anomWWR3}
\end{align}

The $\gamma uu$ and $\gamma \tilde d \tilde d$ couplings are given by 
\begin{align}
{\cal L}^{\gamma u \tilde d} &= -i e \sum_{\ell=0}^\infty  \gamma_{\mu}^{(\ell)} 
 \sum_{n, m }^\infty  \bigg\{ 
\hat Q^{ uu}_{L\, \ell n m} \bar u_L^{(n)} \gamma^\mu u_L^{(m)}
+ \hat Q^{ uu}_{R\, \ell n m} \bar u_R^{(n)} \gamma^\mu u_R^{(m)}  \cr
\noalign{\kern 5pt}
&\hskip 2.cm 
+  \sum_{\tilde d_1, \tilde d_2} \Big( 
\hat Q^{\tilde d_1 \tilde d_2}_{L\, \ell n m} \,  \bar{\tilde d}_{1 L}^{(n)} \gamma^\mu \tilde d_{2L}^{(m)} 
+ \hat Q^{\tilde d_1 \tilde d_2}_{R\, \ell n m} \,  \bar{\tilde d}_{1 R}^{(n)} \gamma^\mu \tilde d_{2R}^{(m)}  \Big) \bigg\} ~, \cr
\noalign{\kern 5pt}
\hat Q^{uu}_{L/R\, \ell n m} &= Q_u \frac{k}{2} \sqrt{kL} \int_{-L}^{L} dy \, e^{\sigma (y)} \, 
 {\cal K}^u_\gamma [ h_{\gamma^{(\ell)}};  ( f, g)^{u^{(n)}}_{L/R}, ( f, g)^{u^{(m)}}_{L/R} ] ~, \cr
\noalign{\kern 5pt}
\hat Q^{\tilde d_1 \tilde d_2}_{L/R\, \ell n m} &= Q_d \frac{k}{2} \sqrt{kL} \int_{-L}^{L} dy \, e^{\sigma (y)} \, 
 {\cal K}^d_\gamma [ h_{\gamma^{(\ell)}};  ( f, g, h, k)^{\tilde d_1^{(n)}}_{L/R}, ( f, g, h, k)^{\tilde d_2^{(m)}}_{L/R} ] ~
\label{gammauu1}
\end{align}
where $e= g_w \sin \theta_W^0$ and $(Q_u , Q_d) = (\frac{2}{3}, - \frac{1}{3})$.  Here
\begin{align}
&{\cal K}^u_\gamma [ h_{\gamma};  ( f, g)_1, ( f, g)_2] = h_\gamma (f_1^* f_2 + g_1^* g_2) ~, \cr
\noalign{\kern 5pt}
&{\cal K}^d_\gamma [ h_{\gamma};  ( f, g, h, k)_1, ( f, g, h, k)_2] 
= h_\gamma (f_1^* f_2 + g_1^* g_2 + h_1^* h_2 + k_1^* k_2) ~ .
\label{gammauu2}
\end{align}
For the photon $\gamma^{(0)}$, $h_{\gamma^{(0)}} = 1/\sqrt{kL}$ and the couplings are diagonal; 
$\hat Q^{uu}_{L/R\, 0 n m} = Q_u \,  \delta_{nm}$ and 
$\hat Q^{\tilde d_1 \tilde d_2}_{L/R\, 0 nm} = Q_d \,  \delta_{nm} \delta_{\tilde d_1 \tilde d_2}$.

The $Zuu$ and $Z \tilde d \tilde d$ couplings are written as
\begin{align}
{\cal L}^{Z u \tilde d} &= -i \frac{g_w}{\cos \theta_W^0} \sum_{\ell=0}^\infty  Z_{\mu}^{(\ell)} 
 \sum_{n, m }^\infty  \bigg\{ 
\hat g^{Zuu}_{L\, \ell n m} \bar u_L^{(n)} \gamma^\mu u_L^{(m)}
+ \hat g^{Zuu}_{R\, \ell n m} \bar u_R^{(n)} \gamma^\mu u_R^{(m)}  \cr
\noalign{\kern 5pt}
&\hskip 2.cm 
+  \sum_{\tilde d_1, \tilde d_2} \Big( 
\hat g^{Z \tilde d_1 \tilde d_2}_{L\, \ell n m} \,  \bar{\tilde d}_{1 L}^{(n)} \gamma^\mu \tilde d_{2L}^{(m)} 
+ \hat g^{Z \tilde d_1 \tilde d_2}_{R\, \ell n m} \,  \bar{\tilde d}_{1 R}^{(n)} \gamma^\mu \tilde d_{2R}^{(m)}  \Big) \bigg\} ~, \cr
\noalign{\kern 5pt}
&\begin{pmatrix} \hat g^{Zuu}_{L/R\, \ell n m}  \cr  \noalign{\kern 5pt}
\hat g^{Z\tilde d_1 \tilde d_2}_{L/R\, \ell n m} \end{pmatrix}
= \begin{pmatrix} \hat g^{Zuu, su2}_{L/R \, \ell n m}  \cr  \noalign{\kern 3pt}
\hat g^{Z\tilde d_1 \tilde d_2, su2}_{L/R \, \ell n m} \end{pmatrix} 
- \sin^2 \theta_W^0 \begin{pmatrix} \hat g^{Zuu, em}_{L/R\, \ell n m} \cr  \noalign{\kern 3pt}
\hat g^{Z\tilde d_1 \tilde d_2, em}_{L/R\, \ell n m} \end{pmatrix}.
\label{Zuu1}
\end{align}
The couplings have been decomposed into the $su2$ and $em$ parts in light of the decomposition in Eq.\ (\ref{Ztower2}).
They are given by
\begin{align}
 \begin{pmatrix} \hat g^{Zuu, su2}_{L/R \, \ell n m}  \cr  \noalign{\kern 3pt}
\hat g^{Z\tilde d_1 \tilde d_2, su2}_{L/R \, \ell n m} \end{pmatrix} 
&= \cos \theta_W^0  \frac{k}{2} \sqrt{kL} \int_{-L}^{L} dy \, e^{\sigma (y)} \, 
\begin{pmatrix}T^3_u  {\cal K}_W [(h^L, h^R,  {\hat h})_{Z^{(\ell)}}^{su2} ;  ( f, g)^{u^{(n)}}_{L/R}, ( f, g)^{u^{(m)}}_{L/R} ] \cr
\noalign{\kern 3pt}
T^3_d {\cal K}_W [(h^L, h^R,  {\hat h})_{Z^{(\ell)}}^{su2} ; ( f, g)^{\tilde d_1^{(n)}}_{L/R}, ( f, g)^{\tilde d_2^{(m)}}_{L/R} ] \end{pmatrix}, \cr
\noalign{\kern 5pt}
\begin{pmatrix} \hat g^{Zuu, em}_{L/R\, \ell n m} \cr  \noalign{\kern 3pt}
\hat g^{Z\tilde d_1 \tilde d_2, em}_{L/R\, \ell n m} \end{pmatrix}
&= \cos \theta_W^0  \frac{k}{2} \sqrt{kL} \int_{-L}^{L} dy \, e^{\sigma (y)} \, 
\begin{pmatrix}Q_u  {\cal K}_\gamma^u [h_{Z^{(\ell)}}^{em} ;  ( f, g)^{u^{(n)}}_{L/R}, ( f, g)^{u^{(m)}}_{L/R} ] \cr
\noalign{\kern 3pt}
Q_d {\cal K}_\gamma^d [h_{Z^{(\ell)}}^{em} ; ( f, g, h, k)^{\tilde d_1^{(n)}}_{L/R}, ( f, g, h, k)^{\tilde d_2^{(m)}}_{L/R} ] \end{pmatrix},
\end{align}
where $(T^3_u, T^3_d) = (\onehalf, - \onehalf)$.
The couplings for $(\ell mn) = (000)$ take values close to those in the SM,
$\hat g^Z_\SM = T^3_L - \sin^2 \theta_W^\SM \, Q_\EM$.

For ${\cal M}^{\mu\nu\lambda}_{Z^{(\ell)} Z^{(r)} }(p,k)$
anomaly terms are decomposed as
\begin{align}
i q_\mu {\cal M}^{\mu\nu\lambda}_{Z^{(\ell)} Z^{(r)} }(p,k)^{ud}
&= -   \frac{1}{4 \pi^2} \, \Big( \frac{g_w}{\cos \theta_W^0} \Big)^2 
\ep^{\nu \lambda \alpha \beta} \, p_\alpha k_\beta \, \Big\{ J^{uu}_{Z^{(\ell)} Z^{(r)}} + J^{dd}_{Z^{(\ell)} Z^{(r)}} \Big\}
~, \cr
\noalign{\kern 5pt}
\begin{pmatrix} J^{uu}_{Z^{(\ell)} Z^{(r)}}  \cr  \noalign{\kern 3pt}  J^{dd}_{Z^{(\ell)} Z^{(r)}} \end{pmatrix} 
&= \begin{pmatrix}
 \Tr \hat g^{Zu u}_{L \, r}  \hat g^{Zuu}_{L \, \ell}  - \Tr \hat g^{Zuu}_{R \, r}  \hat g^{Zuu}_{R \, \ell}  \cr
 \noalign{\kern 3pt}
 \sum_{\tilde d_1, \tilde d_2} \big(  \Tr \hat g^{Z\tilde d_1, \tilde d_2}_{L \, r}  \hat g^{Z\tilde d_2, \tilde d_1}_{L \, \ell}  
 - \Tr \hat g^{Z\tilde d_1, \tilde d_2}_{R \, r}  \hat g^{Z\tilde d_2, \tilde d_1}_{R \, \ell}  \big)  \end{pmatrix}   \cr
\noalign{\kern 5pt}
&= \begin{pmatrix} (T^3_u )^2 F^{1 \, uu}_{Z^{(\ell)} Z^{(r)}}  \cr  \noalign{\kern 3pt}
(T^3_d )^2 F^{1 \, dd}_{Z^{(\ell)} Z^{(r)}}  \end{pmatrix} 
- 2 \sin^2 \theta_W^0  \begin{pmatrix} T^3_u Q_u F^{2 \, uu}_{Z^{(\ell)} Z^{(r)}}  \cr  \noalign{\kern 3pt}
T^3_d Q_d F^{2 \, dd}_{Z^{(\ell)} Z^{(r)}}  \end{pmatrix}    \cr
\noalign{\kern 5pt}
&\hskip 3.5cm
+ \sin^4 \theta_W^0   \begin{pmatrix} (Q_u )^2 F^{3 \, uu}_{Z^{(\ell)} Z^{(r)}}  \cr  \noalign{\kern 3pt}
(Q_d )^2 F^{3 \, dd}_{Z^{(\ell)} Z^{(r)}}  \end{pmatrix} , 
\label{anomZZ1}
\end{align}
where $\big( \hat g^{Zuu}_{L/R \, \ell}  \big)_{nm} = \hat g^{Zuu}_{L/R \, \ell nm}$ etc. 
It is seen by the first method described above for the $F_{W^{(\ell)} W^{\dagger (r)} }^{ud}$ case
that universality relations hold.
\begin{align}
&F^{1 \, uu}_{Z^{(\ell)} Z^{(r)}} = F^{1 \, dd}_{Z^{(\ell)} Z^{(r)}} \equiv F^{1}_{Z^{(\ell)} Z^{(r)}} ~, \cr
\noalign{\kern 5pt}
&F^{2 \, uu}_{Z^{(\ell)} Z^{(r)}} = F^{2 \, dd}_{Z^{(\ell)} Z^{(r)}} \equiv F^{2}_{Z^{(\ell)} Z^{(r)}} ~, \cr
\noalign{\kern 5pt}
&F^{3 \, uu}_{Z^{(\ell)} Z^{(r)}} = F^{3 \, dd}_{Z^{(\ell)} Z^{(r)}} = 0 ~.
\label{anomZZ2}
\end{align}
Further, by the second method, namely by first doing KK sums and then doing the $y$-integral, one finds 
holographic formulas
\begin{align}
F^{1}_{Z^{(\ell)} Z^{(r)}}  &= \frac{kL}{2} \cos^2 \theta_W^0 
\Big\{  h^{L,su2}_{Z^{(\ell)}} h^{L,su2}_{Z^{(r)}}  \big|_{y=0}  - h^{R,su2}_{Z^{(\ell)}}  h^{R,su2}_{Z^{(r)}}  \big|_{y=0}  \cr
\noalign{\kern 5pt}
&\hskip 2.cm 
+ h^{L,su2}_{Z^{(\ell)}} h^{L,su2}_{Z^{(r)}}  \big|_{y=L}  - h^{R,su2}_{Z^{(\ell)}}  h^{R,su2}_{Z^{(r)}}  \big|_{y=L} \Big\}  ~, \cr
\noalign{\kern 5pt}
F^{2}_{Z^{(\ell)} Z^{(r)}}  &= \frac{kL}{4} \cos^2 \theta_W^0 
\Big\{  (h^{L,su2}_{Z^{(\ell)}} - h^{R,su2}_{Z^{(\ell)}} ) h^{em}_{Z^{(r)}}  \big|_{y=0}
+  h^{em}_{Z^{(\ell)}} (  h^{L,su2}_{Z^{(r)}}  -h^{R,su2}_{Z^{(r)}}  )\big|_{y=0}  \cr
\noalign{\kern 5pt}
&\hskip 2.cm 
+  (h^{L,su2}_{Z^{(\ell)}} - h^{R,su2}_{Z^{(\ell)}} ) h^{em}_{Z^{(r)}}  \big|_{y=L}
+  h^{em}_{Z^{(\ell)}} (  h^{L,su2}_{Z^{(r)}}  -h^{R,su2}_{Z^{(r)}}  )\big|_{y=L} \Big\} \cr
\noalign{\kern 5pt}
&= F^{1}_{Z^{(\ell)} Z^{(r)}} ~.
\label{anomZZ3}
\end{align}
In the last equality the relation $ h^{em}_{Z^{(\ell)}}  = h^{L,su2}_{Z^{(\ell)}} +h^{R,su2}_{Z^{(\ell)}} $ has been made use of.
Hence we find that
\begin{align}
i q_\mu {\cal M}^{\mu\nu\lambda}_{Z^{(\ell)} Z^{(r)} }(p,k)^{ud}
&= -   \frac{g_w^2}{8 \pi^2} \,  \big( 1 - 2 \sin^2 \theta_W^0 \big) 
\ep^{\nu \lambda \alpha \beta} \, p_\alpha k_\beta \, 
\frac{1}{\cos^2 \theta_W^0} \, F^{1}_{Z^{(\ell)} Z^{(r)}} ~.
\label{anomZZ4}
\end{align}

For ${\cal M}^{\mu\nu\lambda}_{Z^{(\ell)} \gamma^{(r)} }(p,k)$
anomaly terms are decomposed as
\begin{align}
i q_\mu {\cal M}^{\mu\nu\lambda}_{Z^{(\ell)} \gamma^{(r)} }(p,k)^{ud}
&= -   \frac{1}{4 \pi^2} \,  \frac{g_w \, e}{\cos \theta_W^0} 
\ep^{\nu \lambda \alpha \beta} \, p_\alpha k_\beta \, \Big\{ J^{uu}_{Z^{(\ell)} \gamma^{(r)}} + J^{dd}_{Z^{(\ell)} \gamma^{(r)}} \Big\}
~, \cr
\noalign{\kern 5pt}
\begin{pmatrix} J^{uu}_{Z^{(\ell)} \gamma^{(r)}}  \cr  \noalign{\kern 3pt}  J^{dd}_{Z^{(\ell)} \gamma^{(r)}} \end{pmatrix} 
&= \begin{pmatrix}
 \Tr \hat Q^{u u}_{L \, r}  \hat g^{Zuu}_{L \, \ell}  - \Tr \hat Q^{uu}_{R \, r}  \hat g^{Zuu}_{R \, \ell}  \cr
 \noalign{\kern 3pt}
 \sum_{\tilde d_1, \tilde d_2} \big(  \Tr \hat Q^{\tilde d_1, \tilde d_2}_{L \, r}  \hat g^{Z\tilde d_2, \tilde d_1}_{L \, \ell}  
 - \Tr \hat Q^{\tilde d_1, \tilde d_2}_{R \, r}  \hat g^{Z\tilde d_2, \tilde d_1}_{R \, \ell}  \big)  \end{pmatrix}   \cr
\noalign{\kern 5pt}
&= \begin{pmatrix} Q_u T^3_u  F^{1 \, uu}_{Z^{(\ell)} \gamma^{(r)}}  \cr  \noalign{\kern 3pt}
Q_d T^3_d  F^{1 \, dd}_{Z^{(\ell)} \gamma^{(r)}}  \end{pmatrix} 
- \sin^2 \theta_W^0  \begin{pmatrix}  (Q_u )^2 F^{2 \, uu}_{Z^{(\ell)} \gamma^{(r)}}  \cr  \noalign{\kern 3pt}
(Q_d)^2 F^{2 \, dd}_{Z^{(\ell)} \gamma^{(r)}}  \end{pmatrix}   ,
\label{anomZgamma1}
\end{align}
where $\big( \hat Q^{uu}_{L/R \, \ell}  \big)_{nm} = \hat Q^{uu}_{L/R \, \ell nm}$.
As in the case of ${\cal M}^{\mu\nu\lambda}_{Z^{(\ell)} Z^{(r)} }$ one finds 
\begin{align}
&F^{1 \, uu}_{Z^{(\ell)} \gamma^{(r)}} = F^{1 \, dd}_{Z^{(\ell)} \gamma^{(r)}} \equiv F^{1}_{Z^{(\ell)} \gamma^{(r)}} ~, \cr
\noalign{\kern 5pt}
&F^{2 \, uu}_{Z^{(\ell)} \gamma^{(r)}} = F^{2 \, dd}_{Z^{(\ell)} \gamma^{(r)}} = 0 ~,
\label{anomZgamma2}
\end{align}
and
\begin{align}
F^{1}_{Z^{(\ell)} \gamma^{(r)}}  &= \frac{kL}{2} \cos \theta_W^0 
\Big\{  (h^{L,su2}_{Z^{(\ell)}} - h^{R,su2}_{Z^{(\ell)}} ) h_{\gamma^{(r)}}  \big|_{y=0}
+  (h^{L,su2}_{Z^{(\ell)}} - h^{R,su2}_{Z^{(\ell)}} ) h_{\gamma^{(r)}}  \big|_{y=L} \Big\} ~.
\label{anomZgamma3}
\end{align}
Hence we find that
\begin{align}
i q_\mu {\cal M}^{\mu\nu\lambda}_{Z^{(\ell)} \gamma^{(r)} }(p,k)^{ud}
&= -   \frac{g_w \, e}{8 \pi^2} \,
\ep^{\nu \lambda \alpha \beta} \, p_\alpha k_\beta \, 
\frac{1}{\cos \theta_W^0} \, F^{1}_{Z^{(\ell)} \gamma^{(r)}} ~.
\label{anomZgamma4}
\end{align}
There is no anomaly for ${\cal M}^{\mu\nu\lambda}_{\gamma^{(\ell)} \gamma^{(r)} }$, i.e.\ 
 $i q_\mu {\cal M}^{\mu\nu\lambda}_{\gamma^{(\ell)} \gamma^{(r)} } (p,k) = 0$.

The $Z_R uu$ and $Z_R \tilde d \tilde d$ couplings are given by 
\begin{align}
{\cal L}^{Z_R u \tilde d} &= -i  \frac{g_w}{\cos \theta_W^0}  \sum_{\ell=1}^\infty  Z_{R \mu}^{(\ell)} 
 \sum_{n, m }^\infty  \bigg\{ 
\hat g^{Z_R uu}_{L\, \ell n m} \bar u_L^{(n)} \gamma^\mu u_L^{(m)}
+ \hat g^{Z_R uu}_{R\, \ell n m} \bar u_R^{(n)} \gamma^\mu u_R^{(m)}  \cr
\noalign{\kern 5pt}
&\hskip 2.cm 
+  \sum_{\tilde d_1, \tilde d_2} \Big( 
\hat g^{Z_R \tilde d_1 \tilde d_2}_{L\, \ell n m} \,  \bar{\tilde d}_{1 L}^{(n)} \gamma^\mu \tilde d_{2L}^{(m)} 
+ \hat g^{Z_R \tilde d_1 \tilde d_2}_{R\, \ell n m} \,  \bar{\tilde d}_{1 R}^{(n)} \gamma^\mu \tilde d_{2R}^{(m)}  \Big) \bigg\} ~, \cr
\noalign{\kern 5pt}
\hat g^{Z_R uu}_{L/R\, \ell n m} &= \cos\theta_W^0  \frac{k}{2} \sqrt{kL} \int_{-L}^{L} dy \, e^{\sigma (y)} \, 
\Big\{ T^3_u {\cal K}_W [(h^L. h^R, 0)_{Z_R^{\ell)}}; ( f, g)^{u^{(n)}}_{L/R}, ( f, g)^{u^{(m)}}_{L/R} ]  \cr
\noalign{\kern 5pt}
&\hskip 3.cm
+ (g_B/g_A) Q_X^u   {\cal K}^u_\gamma [ h^B_{{Z_R}^{(\ell)}};  ( f, g)^{u^{(n)}}_{L/R}, ( f, g)^{u^{(m)}}_{L/R} ] \Big\}~, \cr
\noalign{\kern 5pt}
\hat g^{Z_R \tilde d_1 \tilde d_2}_{L/R\, \ell n m} &= \cos\theta_W^0  \frac{k}{2} \sqrt{kL} \int_{-L}^{L} dy \, e^{\sigma (y)} \, 
\Big\{ T^3_d {\cal K}_W [(h^L. h^R, 0)_{Z_R^{\ell)}}; ( f, g)^{ \tilde d_1^{(n)}}_{L/R}, ( f, g)^{\tilde d_2^{(m)}}_{L/R} ]  \cr
\noalign{\kern 5pt}
&\hskip 3.cm
+ (g_B/g_A) \Big( Q_X^d   {\cal K}^u_\gamma [ h^B_{{Z_R}^{(\ell)}};  ( f, g)^{ \tilde d_1^{(n)}}_{L/R}, ( f, g)^{\tilde d_2^{(m)}}_{L/R} ]  \cr
\noalign{\kern 5pt}
&\hskip 4.5cm
+ Q_X^D   {\cal K}^u_\gamma [ h^B_{{Z_R}^{(\ell)}};  ( h, k)^{ \tilde d_1^{(n)}}_{L/R}, ( h, k)^{\tilde d_2^{(m)}}_{L/R} ] \Big) \Big\}~, 
\label{ZRuu1}
\end{align}
where $g_B/g_A = \sin \theta_W^0/\sqrt{1 - 2 \sin^2 \theta_W^0}$, $Q_X^u = Q_X^d = \frac{1}{6}$ and
 $Q_X^D = - \frac{1}{3}$.
Anomaly terms for ${\cal M}^{\mu\nu\lambda}_{Z_R^{(\ell)} Z_R^{(r)} }$ are
\begin{align}
i q_\mu {\cal M}^{\mu\nu\lambda}_{Z_R^{(\ell)} Z_R^{(r)} }(p,k)^{ud}
&= -   \frac{1}{4 \pi^2} \, \Big( \frac{g_w}{\cos \theta_W^0} \Big)^2 
\ep^{\nu \lambda \alpha \beta} \, p_\alpha k_\beta \, \Big\{ J^{uu}_{Z_R^{(\ell)} Z_R^{(r)}} 
+ J^{dd}_{Z_R^{(\ell)} Z_R^{(r)}} \Big\}
~, \cr
\noalign{\kern 5pt}
\begin{pmatrix} J^{uu}_{Z_R^{(\ell)} Z_R^{(r)}}  \cr  \noalign{\kern 3pt}  J^{dd}_{Z_R^{(\ell)} Z_R^{(r)}} \end{pmatrix} 
&= \begin{pmatrix}
 \Tr \hat g^{Z_R u u}_{L \, r}  \hat g^{Z_R uu}_{L \, \ell}  - \Tr \hat g^{Z_R uu}_{R \, r}  \hat g^{Z_R uu}_{R \, \ell}  \cr
 \noalign{\kern 3pt}
 \sum_{\tilde d_1, \tilde d_2} \big(  \Tr \hat g^{Z_R \tilde d_1, \tilde d_2}_{L \, r}  \hat g^{Z_R \tilde d_2, \tilde d_1}_{L \, \ell}  
 - \Tr \hat g^{Z_R \tilde d_1, \tilde d_2}_{R \, r}  \hat g^{Z_R \tilde d_2, \tilde d_1}_{R \, \ell}  \big)  \end{pmatrix}   \cr
\noalign{\kern 5pt}
&= \begin{pmatrix} (T^3_u )^2 F^{1 \, uu}_{Z_R^{(\ell)} Z_R^{(r)}}  \cr  \noalign{\kern 3pt}
(T^3_d )^2 F^{1 \, dd}_{Z_R^{(\ell)} Z_R^{(r)}}  \end{pmatrix} 
+ \frac{2 g_B}{g_A} \begin{pmatrix} T^3_u Q_X^u  F^{2 \, uu}_{Z_R^{(\ell)} Z_R^{(r)}}  \cr  \noalign{\kern 3pt}
T^3_d Q_X^d F^{2 \, dd}_{Z_R^{(\ell)} Z_R^{(r)}}  \end{pmatrix}    \cr
\noalign{\kern 5pt}
&\hskip 3.5cm
+ \frac{g_B^2}{g_A^2}  \begin{pmatrix} (Q_X^u )^2 F^{3 \, uu}_{Z_R^{(\ell)} Z_R^{(r)}}  \cr  \noalign{\kern 3pt}
(Q_X^d )^2 F^{3 \, dd}_{Z_R^{(\ell)} Z_R^{(r)}}  \end{pmatrix} , 
\label{anomZRZR1}
\end{align}
One finds 
\begin{align}
&F^{1 \, uu}_{Z_R^{(\ell)} Z_R^{(r)}} = F^{1 \, dd}_{Z_R^{(\ell)} Z_R^{(r)}} \equiv F^{1}_{Z_R^{(\ell)} Z_R^{(r)}} ~, \cr
\noalign{\kern 5pt}
&F^{2 \, uu}_{Z_R^{(\ell)} Z_R^{(r)}} = F^{2 \, dd}_{Z_R^{(\ell)} Z_R^{(r)}} \equiv F^{2}_{Z_R^{(\ell)} Z_R^{(r)}} ~, \cr
\noalign{\kern 5pt}
&F^{3 \, uu}_{Z_R^{(\ell)} Z_R^{(r)}} = F^{3 \, dd}_{Z_R^{(\ell)} Z_R^{(r)}} = 0 ~.
\label{anomZRZR2}
\end{align}
Noting that 
$h_{Z_R^{(\ell)}}^L + h_{Z_R^{(\ell)}}^R = - (g_B/g_A)  h_{Z_R^{(\ell)}}^B$ and $h_{Z_R^{(\ell)}}^{L/R}(0)= 0$, 
we find that
\begin{align}
F^{1}_{Z_R^{(\ell)} Z_R^{(r)}}  &= \frac{kL}{2} \cos^2 \theta_W^0 
\Big\{  
 h^{L}_{Z_R^{(\ell)}} h^{L}_{Z_R^{(r)}}  \big|_{y=L}  - h^{R}_{Z_R^{(\ell)}}  h^{R}_{Z_R^{(r)}}  \big|_{y=L} \Big\}  ~, \cr
\noalign{\kern 5pt}
\frac{g_B}{g_A} \, F^{2}_{Z_R^{(\ell)} Z_R^{(r)}}   &=  - F^{1}_{Z_R^{(\ell)} Z_R^{(r)}} ~.
\label{anomZRZR3}
\end{align}
Hence we find that 
\begin{align}
i q_\mu {\cal M}^{\mu\nu\lambda}_{Z_R^{(\ell)} Z_R^{(r)} }(p,k)^{ud}
&= -   \frac{g_w^2}{8 \pi^2} \, 
\ep^{\nu \lambda \alpha \beta} \, p_\alpha k_\beta \, 
\frac{1}{\cos^2 \theta_W^0} \, F^{1}_{Z_R^{(\ell)} Z_R^{(r)}} ~.
\label{anomZRZR4}
\end{align}

Anomaly terms for ${\cal M}^{\mu\nu\lambda}_{\gamma^{(\ell)} Z_R^{(r)} }$ are
\begin{align}
i q_\mu {\cal M}^{\mu\nu\lambda}_{\gamma^{(\ell)} Z_R^{(r)} }(p,k)^{ud}
&= -   \frac{1}{4 \pi^2} \,  \frac{e g_w}{\cos \theta_W^0} 
\ep^{\nu \lambda \alpha \beta} \, p_\alpha k_\beta \, \Big\{ J^{uu}_{\gamma^{(\ell)} Z_R^{(r)}} 
+ J^{dd}_{\gamma^{(\ell)} Z_R^{(r)}} \Big\}
~, \cr
\noalign{\kern 5pt}
\begin{pmatrix} J^{uu}_{\gamma^{(\ell)} Z_R^{(r)}}  \cr  \noalign{\kern 3pt}  J^{dd}_{\gamma^{(\ell)} Z_R^{(r)}} \end{pmatrix} 
&= \begin{pmatrix}
 \Tr \hat g^{Z_R u u}_{L \, r}  \hat Q^{ uu}_{L \, \ell}  - \Tr \hat g^{Z_R uu}_{R \, r}  \hat Q^{uu}_{R \, \ell}  \cr
 \noalign{\kern 3pt}
 \sum_{\tilde d_1, \tilde d_2} \big(  \Tr \hat g^{Z_R \tilde d_1, \tilde d_2}_{L \, r}  \hat Q^{\tilde d_2, \tilde d_1}_{L \, \ell}  
 - \Tr \hat g^{Z_R \tilde d_1, \tilde d_2}_{R \, r}  \hat Q^{\tilde d_2, \tilde d_1}_{R \, \ell}  \big)  \end{pmatrix}   \cr
\noalign{\kern 5pt}
&= \begin{pmatrix} T^3_u Q_u F^{1 \, uu}_{\gamma^{(\ell)} Z_R^{(r)}}  \cr  \noalign{\kern 3pt}
T^3_d Q_d F^{1 \, dd}_{\gamma^{(\ell)} Z_R^{(r)}}  \end{pmatrix} 
+ \frac{g_B}{g_A} \begin{pmatrix} Q_X^u Q_u  F^{2 \, uu}_{\gamma^{(\ell)} Z_R^{(r)}}  \cr  \noalign{\kern 3pt}
Q_X^d Q_d F^{2 \, dd}_{\gamma^{(\ell)} Z_R^{(r)}}  \end{pmatrix}   . 
\label{anomgammaZR1}
\end{align}
As in the above we find that 
\begin{align}
&F^{1 \, uu}_{\gamma^{(\ell)} Z_R^{(r)}} = F^{1 \, dd}_{\gamma^{(\ell)} Z_R^{(r)}} \equiv F^{1}_{\gamma^{(\ell)} Z_R^{(r)}} ~, \cr
\noalign{\kern 5pt}
&F^{2 \, uu}_{\gamma^{(\ell)} Z_R^{(r)}} = F^{2 \, dd}_{\gamma^{(\ell)} Z_R^{(r)}} = 0 ~,
\label{anogammaZR2}
\end{align}
and 
\begin{align}
F^{1}_{\gamma^{(\ell)} Z_R^{(r)}}  &= \frac{kL}{2} \cos \theta_W^0 
\Big\{ h_{\gamma^{(\ell)}}\big(  h^{L}_{Z_R^{(r)}}    -  h^{R}_{Z_R^{(r)}}  \big) \Big\}_{y=L} ~.
\label{anomgammaZR3}
\end{align}
Hence we find that
\begin{align}
i q_\mu {\cal M}^{\mu\nu\lambda}_{\gamma^{(\ell)} Z_R^{(r)} }(p,k)^{ud}
&= -   \frac{e \, g_w }{8 \pi^2} \,
\ep^{\nu \lambda \alpha \beta} \, p_\alpha k_\beta \, 
\frac{1}{\cos \theta_W^0} \, F^{1}_{\gamma^{(\ell)} Z_R^{(r)}} ~.
\label{anomgammaZR4}
\end{align}

Finally anomaly terms for ${\cal M}^{\mu\nu\lambda}_{Z^{(\ell)} Z_R^{(r)} }$ are
\begin{align}
i q_\mu {\cal M}^{\mu\nu\lambda}_{Z^{(\ell)} Z_R^{(r)} }(p,k)^{ud}
&= -   \frac{1}{4 \pi^2} \, \Big( \frac{g_w}{\cos \theta_W^0} \Big)^2 
\ep^{\nu \lambda \alpha \beta} \, p_\alpha k_\beta \, \Big\{ J^{uu}_{Z^{(\ell)} Z_R^{(r)}} + J^{dd}_{Z^{(\ell)} Z_R^{(r)}} \Big\} ~, \cr
\noalign{\kern 5pt}
\begin{pmatrix} J^{uu}_{Z^{(\ell)} Z_R^{(r)}}  \cr  \noalign{\kern 3pt}  J^{dd}_{Z^{(\ell)} Z_R^{(r)}} \end{pmatrix} 
&= \begin{pmatrix}
 \Tr \hat g^{Z_R u u}_{L \, r}  \hat g^{Zuu}_{L \, \ell}  - \Tr \hat g^{Z_R uu}_{R \, r}  \hat g^{Zuu}_{R \, \ell}  \cr
 \noalign{\kern 3pt}
 \sum_{\tilde d_1, \tilde d_2} \big(  \Tr \hat g^{Z_R \tilde d_1, \tilde d_2}_{L \, r}  \hat g^{Z\tilde d_2, \tilde d_1}_{L \, \ell}  
 - \Tr \hat g^{Z_R \tilde d_1, \tilde d_2}_{R \, r}  \hat g^{Z\tilde d_2, \tilde d_1}_{R \, \ell}  \big)  \end{pmatrix}   \cr
\noalign{\kern 5pt}
&= \begin{pmatrix} (T^3_u )^2 F^{1 \, uu}_{Z^{(\ell)} Z_R^{(r)}}  \cr  \noalign{\kern 3pt}
(T^3_d )^2 F^{1 \, dd}_{Z^{(\ell)} Z_R^{(r)}}  \end{pmatrix} 
- \sin^2 \theta_W^0  \begin{pmatrix} T^3_u Q_u F^{2 \, uu}_{Z^{(\ell)} Z_R^{(r)}}  \cr  \noalign{\kern 3pt}
T^3_d Q_d F^{2 \, dd}_{Z^{(\ell)} Z_R^{(r)}}  \end{pmatrix}    \cr
\noalign{\kern 5pt}
&
+ \frac{g_B}{g_A} \begin{pmatrix} T^3_u Q_X^u F^{3 \, uu}_{Z^{(\ell)} Z_R^{(r)}}  \cr  \noalign{\kern 3pt}
T^3_d Q_X^d F^{3 \, dd}_{Z^{(\ell)} Z_R^{(r)}}  \end{pmatrix}  
- \sin^2\theta_W^0 \frac{g_B}{g_A} 
  \begin{pmatrix} Q_u Q_X^u  F^{4 \, uu}_{Z^{(\ell)} Z_R^{(r)}}  \cr  \noalign{\kern 3pt}
Q_d Q_X^d F^{4 \, dd}_{Z^{(\ell)} Z_R^{(r)}}  \end{pmatrix} , 
\label{anomZZR1}
\end{align}
We find that
\begin{align}
&F^{j \, uu}_{Z^{(\ell)} Z_R^{(r)}} = F^{j \, dd}_{Z^{(\ell)} Z_R^{(r)}} \equiv F^{j}_{Z^{(\ell)} Z_R^{(r)}} ~,~~ (j=1,2,3)\cr
\noalign{\kern 5pt}
&F^{4 \, uu}_{Z^{(\ell)} Z_R^{(r)}} = F^{4 \, dd}_{Z^{(\ell)} Z_R^{(r)}} = 0 ~,
\label{anomZZR2}
\end{align}
and 
\begin{align}
F^{1}_{Z^{(\ell)} Z_R^{(r)}}  &= \frac{kL}{2} \cos^2 \theta_W^0 
\Big\{  
 h^{L,su2}_{Z^{(\ell)}} h^{L}_{Z_R^{(r)}}  \big|_{y=L}  - h^{R,su2}_{Z^{(\ell)}}  h^{R}_{Z_R^{(r)}}  \big|_{y=L} \Big\}  ~, \cr
\noalign{\kern 5pt}
F^{2}_{Z^{(\ell)} Z_R^{(r)}}  &= \frac{kL}{2} \cos^2 \theta_W^0 
\Big\{    h^{em}_{Z^{(\ell)}} (  h^{L}_{Z_R^{(r)}}  - h^{R}_{Z_R^{(r)}}  ) \Big\}_{y=L} ~,  \cr
\noalign{\kern 5pt}
F^{3}_{Z^{(\ell)} Z_R^{(r)}}  &= \frac{kL}{2} \cos^2 \theta_W^0 
\Big\{   (  h^{L, su2}_{Z^{(\ell)}}  - h^{R, su2}_{Z^{(\ell)}}  )   h^{B}_{Z_R^{(r)}}  \Big\}_{y=L} ~.
\label{anomZZR3}
\end{align}
Hence we see that
\begin{align}
i q_\mu {\cal M}^{\mu\nu\lambda}_{Z^{(\ell)} Z_R^{(r)} }(p,k)^{ud}
&= -   \frac{g_w^2}{8 \pi^2} \,
\ep^{\nu \lambda \alpha \beta} \, p_\alpha k_\beta \, 
\frac{1}{\cos^2 \theta_W^0} \, \Big\{ F^{1}_{Z^{(\ell)} Z_R^{(r)}} -  \sin^2 \theta_W^0 F^{2}_{Z^{(\ell)} Z_R^{(r)}} \Big\} ~.
\label{anomZZR4}
\end{align}

We also note that $A^{\hat 4}_\mu$ field do not contribute to anomalies of the baryon number.
$A^{\hat 4}_\mu$ obeys Dirichlet boundary conditions at both UV and IR branes so that anomaly terms cancel
when contributions from all KK modes of fermions are taken into account.

Numerical values of the $F$ factors for $\theta_H=0.1$ and $m_\KK = 13\,$TeV are given in Appendix C.

\section{5D anomaly formula}

In the previous section we have evaluated coefficients of anomaly terms in $\dd_\mu j_B^\mu$ for all possible 4D gauge fields.
Assembling those results, we are going to derive an anomaly formula for ${J^M_B}_{; M}$ in 5D.
A remarkable fact is that all of the $F$ factors are expressed in terms of values of wave functions of 4D gauge fields
at the UV and IR branes, in a systematic pattern.

Let us first take a look at the anomaly terms of charged vector bosons $W_\mu^{(n)}$ and $W_{R \mu}^{(n)}$.
One generation of quark multiplets gives
\begin{align}
&\int d^4 x \, e^{-iqx}   \la X^{(\ell)} (p) Y^{(r)} (k) | \dd_\mu  j_B^\mu (x) | 0\ra \cr
\noalign{\kern 5pt}
&
= (2\pi)^4  \delta^{(4)} (p + k -q)  \ep_\nu^*(p) \ep_\lambda^* (k)
\, i q_\mu {\cal M}^{\mu\nu\lambda}_{X^{(\ell)} Y^{(r)} }(p,k)      \cr
\noalign{\kern 5pt}
&= (2\pi)^4  \delta^{(4)} (p + k -q)  \Big( - \frac{g_w^2}{8\pi^2} \, \ep^{\nu\lambda \alpha\beta}  
\ep_\nu^*(p) \ep_\lambda^* (k) p_\alpha k_\beta \, F_{X^{(\ell)} Y^{(r)}} \Big)  \cr
\noalign{\kern 5pt}
&= \int d^4 x \, e^{-iqx}   \la X^{(\ell)} (p) Y^{(r)} (k) | \Big\{ - \frac{g_w^2}{16 \pi^2} \, F_{X^{(\ell)} Y^{(r)}}
 X^{(\ell)}_{\mu\nu} \, \tilde Y^{(r) \mu\nu} (x)  \Big\}  | 0\ra , \cr
\noalign{\kern 5pt}
&(X,Y) = (W, W^\dagger), ~ (W_R, W_R^\dagger), ~(W, W_R^\dagger), ~(W_R, W^\dagger) ~, \cr
\noalign{\kern 5pt}
&X_{\mu\nu} = \dd_\mu X_\nu - \dd_\nu X_\mu  ~,~~
\tilde X^{\mu\nu} = \frac{1}{2} \,  \ep^{\mu\nu\alpha\beta} X_{\alpha\beta} ~.
\label{ChargedAnom1}
\end{align}
Here $F$ factors are given by
\begin{align}
&\begin{pmatrix}  F_{W^{(\ell)} W^{\dagger (r)}} \cr \noalign{\kern 3pt}
F_{W_R^{(\ell)} W_R^{\dagger (r)}}  \cr \noalign{\kern 3pt}
F_{W^{(\ell)} W_R^{\dagger (r)}}\end{pmatrix} \cr
\noalign{\kern 5pt}
&= \frac{1}{2} \begin{pmatrix} \big( \bar h_{W^{(r)}}^{L \, *}  \bar h_{W^{(\ell)}}^{L} - \bar h_{W^{(r)}}^{R \, *}  \bar h_{W^{(\ell)}}^{R} \big)_{y=0}
+ \big( \bar h_{W^{(r)}}^{L \, *}  \bar h_{W^{(\ell)}}^{L} - \bar h_{W^{(r)}}^{R \, *}  \bar h_{W^{(\ell)}}^{R} \big)_{y=L}  \cr
\noalign{\kern 3pt}
\big( \bar h_{W_R^{(r)}}^{L \, *}  \bar h_{W_R^{(\ell)}}^{L} - \bar h_{W_R^{(r)}}^{R \, *}  \bar h_{W_R^{(\ell)}}^{R} \big)_{y=L} \cr
\noalign{\kern 3pt}
\big( \bar h_{W_R^{(r)}}^{L \, *}  \bar h_{W^{(\ell)}}^{L} - \bar h_{W_R^{(r)}}^{R \, *}  \bar h_{W^{(\ell)}}^{R} \big)_{y=L} \end{pmatrix}
\label{ChargedAnom2}
\end{align}
where
\begin{align}
\bar h_{W^{(\ell)}}^{L/R} (y) &=  \sqrt{kL} \, h_{W^{(\ell)}}^{L/R} (y) ~, \cr
\noalign{\kern 5pt}
\bar h_{W_R^{(\ell)}}^{L/R} (y) &=  \sqrt{kL} \, h_{W_R^{(\ell)}}^{L/R} (y) ~.
\label{hbar1}
\end{align}
Making use of the KK expansion in Eq.\ (\ref{Wtower1}) and recalling $h_{W_R^{(\ell)}}^{L/R} \big|_{y=0} = 0$, 
one recognizes that the relations (\ref{ChargedAnom1}) and
(\ref{ChargedAnom2}) are summarized as
\begin{align}
\dd_\mu  j_B^\mu (x) \big|_{\rm charged ~bosons}
&= - \frac{g_w^2}{16 \pi^2} \cdot \frac{L}{4} \bigg[ \sum_{a=1}^2 \Big( F^{a_L}_{\mu\nu} \tilde F^{a_L \mu\nu} 
- F^{a_R}_{\mu\nu} \tilde F^{a_R \mu\nu} \Big)_{y=0} \cr
\noalign{\kern 5pt}
&\hskip 2.2cm
+  \sum_{a=1}^2 \Big( F^{a_L}_{\mu\nu} \tilde F^{a_L \mu\nu} 
- F^{a_R}_{\mu\nu} \tilde F^{a_R \mu\nu} \Big)_{y=L}  \bigg] ~.
\label{ChargedAnom3}
\end{align}

For neutral gauge bosons $Z^{(n)}_\mu$,  $Z^{(n)}_{R \mu}$ and $\gamma^{(n)}_\mu$, expressions become
a little more involved as a result of the mixture of $SO(5)$ and $U(1)_X$.  
There is no anomaly for $\gamma^{(\ell)} \gamma^{(r)}$, i.e.\ $
i q_\mu {\cal M}^{\mu\nu\lambda}_{\gamma^{(\ell)} \gamma^{(r)} }(p,k) =0$.  
With the notation for one generation of quark multiplets
\begin{align}
&\int d^4 x \, e^{-iqx}   \la X^{(\ell)} (p) Y^{(r)} (k) | \dd_\mu  j_B^\mu (x) | 0\ra \cr
\noalign{\kern 5pt}
&= \int d^4 x \, e^{-iqx}   \la X^{(\ell)} (p) Y^{(r)} (k) | \Big\{ - \frac{g_w^2}{16 \pi^2 \cos^2 \theta_W^0} \, 
\hat F_{X^{(\ell)} Y^{(r)}} X^{(\ell)}_{\mu\nu} \, \tilde Y^{(r) \mu\nu} (x)  \Big\}  | 0\ra , \cr
\noalign{\kern 5pt}
&\qquad
(X, Y = Z, Z_R, \gamma )~, 
\label{NeutralAnom1}
\end{align}
we derived in the previous section that 
\begin{align}
\hat F_{Z^{(\ell)} Z^{(r)}} &= (1 - 2 \sin^2 \theta_W^0)  F_{Z^{(\ell)} Z^{(r)}}^1 ~,  \cr
\noalign{\kern 5pt}
\hat F_{Z^{(\ell)} \gamma^{(r)}} &= \sin \theta_W^0 \cos \theta_W^0  F_{Z^{(\ell)} \gamma^{(r)}}^1 ~,  \cr
\noalign{\kern 5pt}
\hat F_{Z_R^{(\ell)} Z_R^{(r)}} &=   F_{Z_R^{(\ell)} Z_R^{(r)}}^1 ~,  \cr
\noalign{\kern 5pt}
\hat F_{\gamma^{(\ell)} Z_R^{(r)}} &= \sin \theta_W^0 \cos \theta_W^0  F_{\gamma^{(\ell)} Z_R^{(r)}}^1 ~,  \cr
\noalign{\kern 5pt}
\hat F_{Z^{(\ell)} Z_R^{(r)}} &=   F_{Z^{(\ell)} Z_R^{(r)}}^1  -  \sin^2 \theta_W^0 F_{Z^{(\ell)} Z_R^{(r)}}^2  ~,
\label{NeutralAnom2}
\end{align}
where
\begin{align}
F_{Z^{(\ell)} Z^{(r)}}^1 &= \frac{1}{2} 
\Big\{  (\bar h^{L,su2}_{Z^{(\ell)}} \bar h^{L,su2}_{Z^{(r)}}  - \bar h^{R,su2}_{Z^{(\ell)}}  \bar h^{R,su2}_{Z^{(r)}})_{y=0} 
+  (\bar h^{L,su2}_{Z^{(\ell)}} \bar h^{L,su2}_{Z^{(r)}}  - \bar h^{R,su2}_{Z^{(\ell)}}  \bar h^{R,su2}_{Z^{(r)}})_{y=L} \Big\}, \cr
\noalign{\kern 5pt}
F_{Z^{(\ell)} \gamma^{(r)}}^1 &= \frac{1}{2}  
\Big\{  (\bar h^{L,su2}_{Z^{(\ell)}} - \bar h^{R,su2}_{Z^{(\ell)}} ) \bar h_{\gamma^{(r)}}  \big|_{y=0}
+  (\bar h^{L,su2}_{Z^{(\ell)}} - \bar h^{R,su2}_{Z^{(\ell)}} ) \bar h_{\gamma^{(r)}}  \big|_{y=L} \Big\} , \cr
\noalign{\kern 5pt}
F_{Z_R^{(\ell)} Z_R^{(r)}}^1 &= \frac{1}{2} \,
\big(  \bar h^{L}_{Z_R^{(\ell)}} \bar h^{L}_{Z_R^{(r)}}  - \bar h^{R}_{Z_R^{(\ell)}}  \bar h^{R}_{Z_R^{(r)}} \big)_{y=L} , \cr
\noalign{\kern 5pt}
F_{\gamma^{(\ell)} Z_R^{(r)}}^1 &=  \frac{1}{2}  \,
 \bar h_{\gamma^{(\ell)}}\big(  \bar h^{L}_{Z_R^{(r)}}    -  \bar h^{R}_{Z_R^{(r)}}  \big)  \big|_{y=L}, \cr
\noalign{\kern 5pt}
F_{Z^{(\ell)} Z_R^{(r)}}^1 &= \frac{1}{2} \,
\big( \bar h^{L,su2}_{Z^{(\ell)}} \bar h^{L}_{Z_R^{(r)}}   - \bar h^{R,su2}_{Z^{(\ell)}}  \bar h^{R}_{Z_R^{(r)}}  \big)_{y=L} ~, \cr
\noalign{\kern 5pt}
F_{Z^{(\ell)} Z_R^{(r)}}^2 &= \frac{1}{2} \,
\bar  h^{em}_{Z^{(\ell)}} ( \bar h^{L}_{Z_R^{(r)}}  - \bar h^{R}_{Z_R^{(r)}}  ) \big|_{y=L} .
\label{NeutralAnom3}
\end{align}
and
\begin{align}
\bar h^{L/R,su2}_{Z^{(\ell)}} (y) &= \sqrt{kL} \cos \theta_W^0 \, h^{L/R,su2}_{Z^{(\ell)}}  (y) ~, \cr
\noalign{\kern 5pt}
\bar  h^{em}_{Z^{(\ell)}} (y) &= \sqrt{kL} \cos \theta_W^0 \,  h^{em}_{Z^{(\ell)}} (y) ~, \cr
\noalign{\kern 5pt}
\bar h_{\gamma^{(\ell)}} (y) &= \sqrt{kL} \,h_{\gamma^{(\ell)}}  (y) ~, \cr
\noalign{\kern 5pt}
\bar h^{L/R}_{Z_R^{(\ell)}} (y) &= \sqrt{kL} \cos \theta_W^0 \, h^{L/R}_{Z_R^{(\ell)}}  (y) ~.
\label{hbar2}
\end{align}
Making use of the KK expansion in Eqs.\ (\ref{Ztower1}) and (\ref{Ztower2}) and recalling 
$h_{Z^{(\ell)}}^{em} = h_{Z^{(\ell)}}^{L, su2} + h_{Z^{(\ell)}}^{R, su2}$ and 
$h_{Z_R^{(\ell)}}^{L/R} \big|_{y=0} = 0$, 
the above relations (\ref{NeutralAnom1}),  (\ref{NeutralAnom2})  and (\ref{NeutralAnom3}) 
are summarized as
\begin{align}
\dd_\mu  j_B^\mu (x) \big|_{\rm neutral ~bosons}
&= - \frac{g_w^2}{16 \pi^2} \cdot \frac{L}{4} \bigg[ \Big( F^{3_L}_{\mu\nu} \tilde F^{3_L \mu\nu} 
- F^{3_R}_{\mu\nu} \tilde F^{3_R \mu\nu} \Big)_{y=0} \cr
\noalign{\kern 5pt}
&\hskip 2.2cm
+  \Big( F^{3_L}_{\mu\nu} \tilde F^{3_L \mu\nu} 
- F^{3_R}_{\mu\nu} \tilde F^{3_R \mu\nu} \Big)_{y=L}  \bigg] ~.
\label{NeutralAnom4}
\end{align}

Combining Eqs.\ (\ref{ChargedAnom3}) and (\ref{NeutralAnom4}) and multiplying a generation factor $N_f=3$,
one obtains
\begin{align}
\dd_\mu  j_B^\mu (x) 
&= -  \frac{N_f g_w^2}{16 \pi^2} \cdot \frac{L}{4} \bigg[ \sum_{a=1}^3 \Big( F^{a_L}_{\mu\nu} \tilde F^{a_L \mu\nu} 
- F^{a_R}_{\mu\nu} \tilde F^{a_R \mu\nu} \Big)_{y=0} \cr
\noalign{\kern 5pt}
&\hskip 2.2cm
+  \sum_{a=1}^3 \Big( F^{a_L}_{\mu\nu} \tilde F^{a_L \mu\nu} 
- F^{a_R}_{\mu\nu} \tilde F^{a_R \mu\nu} \Big)_{y=L}  \bigg] ~.
\label{totalAnom1}
\end{align}
Here  $F_{\mu\nu}^{a_{L/R}} (x,y)$ are field strengths in 5D.
Retaining only zero modes $W_\mu = W_\mu^{(0)}$, $Z_\mu = Z_\mu^{(0)}$ and $A^\gamma_\mu = \gamma_\mu^{(0)}$,
one finds
\begin{align}
&\dd_\mu  j_B^\mu (x) \big|_{W,Z,\gamma}
= -  \frac{N_f g_w^2}{16 \pi^2} \bigg\{  F_{W^{(0)} W^{\dagger (0)}}  W_{\mu\nu} \tilde W^{\dagger \mu\nu}  \cr
\noalign{\kern 5pt}
&\hskip 0.5cm
+  F^1_{Z^{(0)} Z^{(0)}} \, \frac{1 - 2 \sin^2 \theta_W^0}{2 \cos^2 \theta_W^0} \,  Z_{\mu\nu} \tilde Z^{\mu\nu}
+  F^1_{Z^{(0)} \gamma^{(0)}} \, \frac{ \sin \theta_W^0}{\cos \theta_W^0} \,  A^\gamma_{\mu\nu} \tilde Z^{\mu\nu} \bigg\} ~. 
\label{totalAnom2}
\end{align}
The values of the $F$ factors for $\theta_H=0.1$ and $m_\KK = 13\,$TeV are 
$F_{W^{(0)} W^{\dagger (0)}} = 0.997728$, $F^1_{Z^{(0)} Z^{(0)}} = 0.997796$ and $F^1_{Z^{(0)} \gamma^{(0)}}  = 0.997649$.
The values of these $F$ factors differ from each other as the wave functions of $W^{(0)}$ and $Z^{(0)}$ for $\theta_H \not= 0$ 
are deformed in a slightly different manner  in the RS space.  
In the $\theta_H \go 0$ limit, in which the  $SU(2)_L \times U(1)_Y$ symmetry is unbroken, 
one finds that $F_{W^{(0)} W^{\dagger (0)}} = F^1_{Z^{(0)} Z^{(0)}} = F^1_{Z^{(0)} \gamma^{(0)}}  =  1$.
The expression (\ref{totalAnom2}) should be compared with the result 
in the SM \cite{Manton1984, Donoghue2014, Smith2026}
\begin{align}
\dd_\mu  j_B^\mu (x) \big|^\SM
&= -  \frac{N_f g_w^2}{16 \pi^2} \bigg\{   W_{\mu\nu} \tilde W^{\dagger \mu\nu}  
+  \frac{1 - 2 s_W^2}{2 c_W^2} \,  Z_{\mu\nu} \tilde Z^{\mu\nu}  
+  \frac{ s_W}{c_W} \,  A^\gamma_{\mu\nu} \tilde Z^{\mu\nu} \bigg\} \cr
\noalign{\kern 5pt}
&  =-  \frac{N_f}{32 \pi^2} \Big\{ g_w^2 \sum_{a=1}^3 W^a_{\mu\nu} \tilde W^{a\, \mu\nu} 
- {g'}^2 B_{\mu\nu} \tilde B^{\mu\nu} \Big\}
\label{totalAnomSM}
\end{align}
where $s_W = \sin \theta_W^\SM$, $c_W = \cos\theta_W^\SM$ and $W^a_\mu (x)$'s 
are  $SU(2)_L$ gauge fields.

Some of the $F$ factors for KK excited modes of gauge fields become large as the values of
their wave functions at the IR brane are large.    
For instance $F_{W^{(1)} W^{\dagger (1)}} = 26.720$.  (See appendix C.)
Although $W^{(1)}$ is heavy ($m_{W^{(1)}} = 10.2\,$TeV), the anomaly term may play an important role
in the left-right phase transition around $T=2.5\,$TeV in the early universe shown in ref.\ \cite{phaseT2021}.

The anomaly equation for the 5D Noether current $J_B^M (x,y)$ becomes
\begin{align}
{J_B^M}_{; M} &= - \frac{N_f g_A^2}{32 \pi^2} \frac{1}{\sqrt{-G}} \big\{ \delta_{2L}(y) + \delta_{2L}(y - L) \big\} 
\sum_{a=1}^3 \Big( F^{a_L}_{\mu\nu} \tilde F^{a_L \mu\nu}  - F^{a_R}_{\mu\nu} \tilde F^{a_R \mu\nu} \Big) .
\label{5DAnomaly1}
\end{align}
Note $g_A = g_w \sqrt{L}$.
Years ago Arkani-Hamed, Cohen and Georgi examined a chiral $U(1)$ gauge theory on $M^4 \times (S^1/Z_2)$ orbifold
with one Weyl fermion \cite{Arkani-Hamed2001} and derived 
\begin{align}
\dd_C J^C (x,x_4) &= \frac{g^2}{32 \pi^2} \big\{ \delta (x_4) + \delta (x_4 - L) \big\} F_{\mu\nu} \tilde F^{\mu\nu} ~.
\label{ACG1}
\end{align}
The formula (\ref{5DAnomaly1}) takes the same form as (\ref{ACG1}).
In the GUT-inspired $SO(5) \times U(1)_X \times SU(3)_C$ GHU in the RS space the anomaly term is proportional
to the difference between $\sum F^{a_L}_{\mu\nu} \tilde F^{a_L \mu\nu} $ in $SU(2)_L$ and 
$\sum F^{a_R}_{\mu\nu} \tilde F^{a_R \mu\nu} $ in $SU(2)_R$.
Eq.\ (\ref{5DAnomaly1}) is gauge-invariant under $SO(4) \simeq SU(2)_L \times SU(2)_R$. 

Anomaly for the lepton number current ${J_L^M}_{;M}$ is analyzed along the same line as for ${J_B^M}_{;M}$.
Universality relations for the various $F$ factors remain true.  For instance the relation (\ref{anomWW3}) 
is extended to $F^\alpha_{W^{(\ell)} W^{\dagger (r)}} = F_{W^{(\ell)} W^{\dagger (r)}}$  
($\alpha = \nu_e e, \nu_\mu \mu, \nu_\tau \tau$).
Consequently one finds that ${\cal F}_L^{\rm anom} = {\cal F}_B^{\rm anom}$ and 
\begin{align}
\big( J_B^M- J_L^M \big)_{;M} &= \delta_{2L} (y) \, {\cal K}_{\rm break} ~,
\label{BminusL}
\end{align}
where ${\cal K}_{\rm break}$ is given in Eq.\ (\ref{LeptonAnom1}).
$B-L$ is conserved up to the ${\cal K}_{\rm break}$ term.

Although it is difficult to construct a realistic GHU model in flat space, the anomaly formula 
(\ref{5DAnomaly1}) itself remains valid in the flat-space limit $k \go 0, z_L \go 1$ with $L$ kept fixed.
In this limit the RS space becomes  $M^4 \times (S^1/Z_2)$ orbifold.
As shown in detail in an $SU(2)$  GHU model in ref.\ \cite{AnomalyFlow2}, $\bar h$ functions have
well-defined limits, and anomalies show behavior of step functions as $\theta_H$ varies.
Singular behavior appears when level crossing of mass spectra occurs, namely at 
$\theta_H = 0, \pm \onehalf \pi$ and $\pi$.

\section{Summary}

Baryon number conservation is violated by anomaly at the quantum level. 
In the SM the anomaly term is summarized as in Eq.\ (\ref{totalAnomSM}).
In the present paper we analyzed baryon number conservation in the GUT-inspired 
$SO(5) \times U(1)_X \times SU(3)_C$ GHU in the RS space, and obtained 
the formulas Eqs.\ (\ref{totalAnom1}) and  (\ref{5DAnomaly1}) in four  and  five dimensions, respectively.  
The anomaly term contains KK modes of gauge fields.
The anomaly coefficients are given by various $F$ factors  in Eqs.\ (\ref{ChargedAnom2})
and (\ref{NeutralAnom3}), which are expressed in terms of the values of wave functions of gauge fields
at the UV and IR branes.  It may be phrased that chiral anomalies appear at the fixed branes in 5D orbifolds.

Physical consequences of the anomaly term need to be explored further.
In 1984 Klinkhamer and Manton found a sphaleron solution in the SM \cite{Manton1984}, which may play an important role 
in baryon number generation in the early universe through the anomaly equation.
A detailed analysis of a sphaleron solution in the GUT-inspired GHU is necessary.  We will come back
to this issue in the near future.

\section*{Data availability}

There are no publicly available research data or software supporting this manuscript.  
Requests for further information or data should be sent to the author.
 
\appendix

\section{Bessel functions in the RS space} 
 
 In the RS space KK mass spectrum and wave functions of KK modes are expressed 
 in terms of special combinations of  Bessel functions.
Basis functions for gauge fields are
\begin{align}
 F_{\alpha, \beta}(u, v) &\equiv J_\alpha(u) Y_\beta(v) - Y_\alpha(u) J_\beta(v) ~, \cr
\noalign{\kern 5pt}
 C(z; \lambda) &= \frac{\pi}{2} \lambda z z_L F_{1,0}(\lambda z, \lambda z_L) ~,  \cr
 S(z; \lambda) &= -\frac{\pi}{2} \lambda  z F_{1,1}(\lambda z, \lambda z_L) ~, \cr
 C^\prime (z; \lambda) &= \frac{\pi}{2} \lambda^2 z z_L F_{0,0}(\lambda z, \lambda z_L) ~,  \cr
S^\prime (z; \lambda) &= -\frac{\pi}{2} \lambda^2 z  F_{0,1}(\lambda z, \lambda z_L)~,  \cr
\check S(z; \lambda) &= \frac{C(1; \lambda)}{S(1; \lambda)} \,  S(z; \lambda) ~, 
\label{functionA1}
\end{align}
where $J_\alpha (u)$ and $Y_\alpha (u)$ are Bessel functions of  the first and second kind.
$C(z;\lambda)$ and $S(z; \lambda)$  satisfy
\begin{align}
&- z \frac{d}{dz} \frac{1}{z} \frac{d}{dz} \begin{pmatrix} C \cr S \end{pmatrix} 
= \lambda^{2} \begin{pmatrix} C \cr S \end{pmatrix} ~.
\label{relationA1}
\end{align}
These functions satisfy boundary conditions  $C = z_L$, $C'  =S = 0 $, and $S' = \lambda$ at $z=z_L$.
A relation $CS' - S C' = \lambda z$ holds.

 For fermion fields with a bulk mass parameter $c$, basis functions are
\begin{align}
\begin{pmatrix} C_L \cr S_L \end{pmatrix} (z; \lambda,c)
&= \pm \frac{\pi}{2} \lambda \sqrt{z z_L} F_{c+\frac12, c\mp\frac12}(\lambda z, \lambda z_L) ~, \cr
\begin{pmatrix} C_R \cr S_R \end{pmatrix} (z; \lambda,c)
&= \mp \frac{\pi}{2} \lambda \sqrt{z z_L} F_{c- \frac12, c\pm\frac12}(\lambda z, \lambda z_L) ~, \cr
\begin{pmatrix} \check S_L \cr \check C_R \end{pmatrix} (z; \lambda,c)
&=\frac{C_L (1; \lambda, c)}{S_L (1; \lambda, c)} \begin{pmatrix} S_L \cr C_R \end{pmatrix} (z; \lambda,c) ~.
\label{functionA2}
\end{align}
These functions satisfy 
\begin{align}
&D_{+} (c) \begin{pmatrix} C_{L} \cr S_{L} \end{pmatrix} = \lambda  \begin{pmatrix} S_{R} \cr C_{R} \end{pmatrix}, \cr
\noalign{\kern 5pt}
&D_{-} (c) \begin{pmatrix} S_{R} \cr C_{R} \end{pmatrix} = \lambda  \begin{pmatrix} C_{L} \cr S_{L} \end{pmatrix}, \cr
\noalign{\kern 5pt}
&D_{\pm} (c) = \pm \frac{d}{dz} + \frac{c}{z} ~, 
\label{relationA2}
\end{align}
with the boundary conditions $C_{R/L} =1$, $D_-(c) C_R = D_+(c) C_L =S_{R/L} = 0$ at $z=z_{L} $, and 
$C_L C_R - S_L S_R=1$.
\ignore{For fermion fields with a vector-like mass $m = k \tilde m$,  mode functions are expressed
in terms of
\begin{align}
{\cal C}_{L/R\, 1}(z; \lambda, c, \tilde m) &= C_{L/R} (z; \lambda, c+\tilde{m})+C_{L/R} (z; \lambda, c-\tilde{m}) ~, \cr
{\cal S}_{L/R\, 1}(z; \lambda, c, \tilde m) &= S_{L/R} (z; \lambda, c+\tilde{m})+S_{L/R} (z; \lambda,c-\tilde{m}) ~, \cr
{\cal C}_{L/R\, 2}(z; \lambda, c, \tilde m) &= S_{L/R} (z; \lambda, c+\tilde{m})-S_{L/R} (z; \lambda, c-\tilde{m}) ~, \cr
{\cal S}_{L/R\, 2}(z; \lambda, c, \tilde m) &= C_{L/R} (z; \lambda, c+\tilde{m})-C_{L/R} (z; \lambda, c-\tilde{m}) ~.
\label{MassiveFermion1}
\end{align}
}
 
\section{Wave functions in the RS space} 
 
Wave functions of all KK modes of gauge bosons and fermions are summarized in Ref.\ \cite{YHbook}.
The $F$ factors for the anomaly coefficients are expressed in terms of the values of wave functions of gauge fields
at the UV brane (at $y=0, z=1$) and at the IR brane (at $y=L, z=z_L$), as shown in Sections 4 and 5.
In this appendix we list  wave functions of gauge fields.

The spectrum $m_{W^{(n)}} = k \lambda_{W^{(n)}}$ ($n \ge 0$) of the $W$ boson tower is determined by 
\begin{align}
&2 S C' (1; \lambda_{W^{(n)}}) +   \lambda_{W^{(n)}} \sin^2 \theta_H  =0 ~.
\label{Wspectrum1}
\end{align}
Wave functions in the $z$-coordinate ($1 \le z \le z_L$) are given by
\begin{align}
\begin{pmatrix}  h^L_{W^{(n)}} (z) \cr \mysnoalign h^R_{W^{(n)}} (z) \end{pmatrix} 
&=  \frac{1 \pm \cos \theta (z)}{2}  \,  \tilde h^L_{W^{(n)}} (z) 
+  \frac{1 \mp \cos \theta (z)}{2}  \,  \tilde h^R_{W^{(n)}} (z) 
\pm \frac{\sin \theta(z)}{\sqrt{2}} \, \tilde{\hat h}_{W^{(n)}} (z) ~, \cr
\noalign{\kern 5pt}
{\hat h}_{W^{(n)}} (z)  &= -  \frac{\sin \theta(z)}{\sqrt{2}} \,  \tilde h^L_{W^{(n)}} (z) 
+ \frac{\sin \theta(z)}{\sqrt{2}} \,  \tilde h^R_{W^{(n)}} (z)  + \cos \theta (z) \, \tilde{\hat h}_{W^{(n)}} (z) ~.
\label{WbosonWave2}
\end{align}
where
\begin{align}
&\theta (z) = \theta_H \, \frac{z_L^2 - z^2}{z_L^2 - 1} ~,  \cr
\noalign{\kern 5pt}
& \begin{pmatrix}  \tilde h^L_{W^{(n)}} (z) \cr \mysnoalign
 \tilde h^R_{W^{(n)}} (z) \cr  \mysnoalign  \tilde{\hat h}_{W^{(n)}} (z) \end{pmatrix} =
 \frac{1}{\sqrt{2 \, r_{W^{(n)}}}}
\begin{pmatrix} (1 + c_H) \, C(z; \lambda_{W^{(n)}}) \cr 
(1 - c_H) \, C(z; \lambda_{W^{(n)}})  \cr  
\sqrt{2} \, s_H \check S(z; \lambda_{W^{(n)}}) \end{pmatrix}  , \cr
\noalign{\kern 5pt}
&\int_1^{z_L}  \frac{dz}{z} \Big\{ ( |\tilde h^L_{W^{(n)}} |^2 + |\tilde h^R_{W^{(n)}} |^2 
+ | \tilde{\hat h}_{W^{(n)}} |^2 \Big\} = 1 ~, \cr
\noalign{\kern 5pt}
&\quad c_H = \cos \theta_H ~, ~~ s_H = \sin \theta_H ~.
\label{WbosonWave1}
\end{align}
Wave functions in the $y$-coordinate, used in Eq.\ (\ref{Wtower1}), are obtained by inserting $z=e^{ky}$ into 
Eq.\ (\ref{WbosonWave2}) in the region $0 \le y \le L$ and extending to other region by the orbifold parity condition 
Eq.\ (\ref{BC-gauge1}). 

The spectrum $m_{W_R^{(n)}} = k \lambda_{W_R^{(n)}}$ ($n \ge 1$) of the $W_R$ boson tower is determined by 
$C(1; \lambda_{W_R^{(n)}}) = 0$.  Wave functions in the $z$-coordinate ($1 \le z \le z_L$) are given by
\begin{align}
\begin{pmatrix}  h^L_{W_R^{(n)}} (z) \cr \mysnoalign h^R_{W_R^{(n)}} (z)   \end{pmatrix} 
& =  \frac{1 \pm \cos \theta (z)}{2}  \,  \tilde h^L_{W_R^{(n)}} (z) 
+  \frac{1 \mp \cos \theta (z)}{2}  \,  \tilde h^R_{W_R^{(n)}} (z) ~, \cr
\noalign{\kern 5pt}
\hat h_{W_R^{(n)}} (z)  &= \frac{\sin \theta (z)}{\sqrt{2}}  \Big\{ -  \tilde h^L_{W_R^{(n)}} (z)  + \tilde h^R_{W_R^{(n)}} (z) \Big\} ,  \cr
\noalign{\kern 5pt}
\begin{pmatrix}  \tilde h^L_{W_R^{(n)}} (z) \cr \mysnoalign  \tilde h^R_{W_R^{(n)}} (z) \end{pmatrix} 
&= \frac{1}{\sqrt{2 \, r_{W_R^{(n)}}}}
\begin{pmatrix} 1 - c_H \cr  -1 - c_H \end{pmatrix}   C(z; \lambda_{W_R^{(n)}})  ~, \cr
\noalign{\kern 5pt}
&\int_1^{z_L}  \frac{dz}{z} \Big\{ ( |\tilde h^L_{W_R^{(n)}} |^2 + |\tilde h^R_{W_R^{(n)}} |^2  \Big\} = 1 ~.
\label{WRbosonWave1}
\end{align}
Wave functions in the $y$-coordinate are obtained in the same manner as for $W^{(n)}$.

The spectrum $m_{\gamma^{(n)}} = k \lambda_{\gamma^{(n)}}$ ($n \ge 0$) of the photon ($\gamma$) tower is 
given by $\lambda_{\gamma^{(0)}} = 0$  and $C' (1; \lambda_{\gamma^{(n)}} )=0$ ($n \ge 1$).
Wave functions in Eq.\ (\ref{Ztower1}) are given, in the $z$-coordinate, by
\begin{align}
&h_{\gamma^{(n)}} (z) = 
\begin{cases} \myfrac{1}{\sqrt{kL}} &{\rm for ~} n=0 , \cr
\myfrac{1}{\sqrt{r_{\gamma^{(n)}}}} \, C(z; \lambda_{\gamma^{(n)}} ) &{\rm for ~} n \ge 1 , \end{cases} \cr
\noalign{\kern 5pt}
&\int_1^{z_L}  \frac{dz}{z} \,  | h_{\gamma^{(n)}} |^2   = 1 ~.
\label{gammaWave1}
\end{align}

The spectrum $m_{Z^{(n)}} = k \lambda_{Z^{(n)}}$ ($n \ge 0$) of the $Z$ boson tower is determined by 
\begin{align}
&2 S C' (1; \lambda_{Z^{(n)}}) +  \lambda_{Z^{(n)}} \frac{\sin^2 \theta_H}{\cos^2 \theta_W^0}=0 ~, \cr
\noalign{\kern 5pt}
&\sin \theta_W^0  = \frac{g_B}{\sqrt{g_A^2 + 2 g_B^2}} ~.
\label{Zspectrum1}
\end{align}
With the decomposition in Eq.\ (\ref{Ztower2}) wave functions in the $z$-coordinate are given by
\begin{align}
\begin{pmatrix}  h^{L, su2}_{Z^{(n)}} (z) \cr \mysnoalign h^{R, su2}_{Z^{(n)}} (z) \end{pmatrix} 
&=  \frac{1 \pm \cos \theta (z)}{2}  \,  \tilde h^{L, su2}_{Z^{(n)}}  (z) 
+  \frac{1 \mp \cos \theta (z)}{2}  \,  \tilde h^{R, su2}_{Z^{(n)}} (z) 
\pm \frac{\sin \theta(z)}{\sqrt{2}} \, \tilde{\hat h}^{su2}_{Z^{(n)}} (z) ~, \cr
\noalign{\kern 5pt}
{\hat h}^{su2}_{Z^{(n)}} (z)  &= -  \frac{\sin \theta(z)}{\sqrt{2}} \,  \tilde h^{L, su2}_{Z^{(n)}}  (z) 
+ \frac{\sin \theta(z)}{\sqrt{2}} \,  \tilde h^{R, su2}_{Z^{(n)}} (z)  + \cos \theta (z) \, \tilde{\hat h}^{su2}_{Z^{(n)}} (z) ~, \cr
\noalign{\kern 5pt}
h^{em}_{Z^{(n)}} (z)  &= \sqrt{\frac{2}{r_{Z^{(n)}}}} \, C(z, \lambda_{Z^{(n)}} )~, 
\label{ZbosonWave3}
\end{align}
where
\begin{align}
&\begin{pmatrix} \tilde h^{L, su2}_{Z^{(n)}} (z) \cr \mysnoalign \tilde h^{R, su2}_{Z^{(n)}} (z) \cr  
\mysnoalign    \tilde {\hat h}^{su2}_{Z^{(n)}} (z)  \end{pmatrix} 
= \frac{1}{\sqrt{ \, 2 \, r_{Z^{(n)}} }} 
\begin{pmatrix} (1 + c_H) \,C(z, \lambda_{Z^{(n)}} ) \cr  (1 -c_H) \,C(z, \lambda_{Z^{(n)}} ) \cr  
  \sqrt{2}s_H  \check S (z, \lambda_{Z^{(n)}}  ) \end{pmatrix} , \cr
\noalign{\kern 5pt}
&\int_1^{z_L}  \frac{dz}{z} \Big\{ ( | h^L_{Z^{(n)}} |^2 + | h^R_{Z^{(n)}} |^2 + |  {\hat h}_{Z^{(n)}} |^2  
+  |h^B_{Z^{(n)}} |^2 \Big\} = 1 ~.
\label{ZbosonWave1}
\end{align}
Note that $h^{L, su2}_{Z^{(n)}} (z)  + h^{R, su2}_{Z^{(n)}} (z)  = h^{em}_{Z^{(n)}} (z) $.

The spectrum $m_{Z_R^{(n)}} = k \lambda_{Z_R^{(n)}}$ ($n \ge 1$) of the $Z_R$ boson tower is the same
as that of the $W_R$ boson tower, being determined by $C(1; \lambda_{Z_R^{(n)}}) = 0$.
Wave function in the $z$-coordinate are given by
\begin{align}
\begin{pmatrix}  h^L_{Z_R^{(n)}} (z) \cr \mysnoalign h^R_{Z_R^{(n)}} (z) \end{pmatrix} 
&=  \frac{1 \pm \cos \theta (z)}{2}  \,  \tilde h^L_{Z_R^{(n)}} (z) 
+  \frac{1 \mp \cos \theta (z)}{2}  \,  \tilde h^R_{Z_R^{(n)}} (z) ~, \cr
\noalign{\kern 5pt}
\hat h_{Z_R^{(n)}} (z)  &= \frac{\sin \theta (z)}{\sqrt{2}}  \Big\{ -  \tilde h^L_{Z_R^{(n)}} (z)  + \tilde h^R_{Z_R^{(n)}} (z) \Big\} ,  \cr
\noalign{\kern 5pt}
h^B_{Z_R^{(n)}} (z)  &=\sqrt{ \frac{2}{r_{Z_R^{(n)}}} }
\frac{c_H \sin \theta_W^0}{\sqrt{1 - 2 \sin^2 \theta_W^0}} \, C(z; \lambda_{Z_R^{(n)}})  ~,  
\label{ZRbosonWave1}
\end{align}
where
\begin{align}
\begin{pmatrix}  \tilde h^L_{Z_R^{(n)}} (z) \cr \mysnoalign  \tilde h^R_{Z_R^{(n)}} (z) \end{pmatrix} 
&= \frac{1}{\sqrt{2 \, r_{Z_R^{(n)}}}}
\begin{pmatrix} 1 - c_H  \cr  -1 - c_H \end{pmatrix}  \, C(z; \lambda_{Z_R^{(n)}})   , \cr
\noalign{\kern 5pt}
r_{Z_R^{(n)}} &=
\int_1^{z_L}  \frac{dz}{z} \bigg\{ 1 + \frac{c_H^2}{1 - 2 \sin^2 \theta_W^0}  \bigg\} \, C(z; \lambda_{Z_R^{(n)}})^2  ~.
\label{ZRbosonWave2}
\end{align}


\section{$F$ factors}

\subsection{Mass spectrum of gauge fields}

Mass spectrum of 4D gauge fields for $\theta_H = 0.1$ and $m_\KK = 13\,$TeV at the tree level is
tabulated in Table \ref{Table:mass}.  
The value of $m_{Z^{(0)}} = 91.1876\,$GeV is an input.  
$z_L = 3.83182 \times 10^{11}$, $kL = 26.6718$, and $\sin^2 \theta_W^0 = 0.230352$.
There is no anomaly associated with $\{ A^{\hat 4 \, (\ell)}_\mu \}$, as $A^{\hat 4}_\mu = 0$ at $y=0$ and $L$.

\def\GeV{$\,${\rm GeV}}
\def\TeV{$\,${\rm TeV}}

\begin{table}[tbh]
\renewcommand{\arraystretch}{1.2}
\begin{center}
\caption{Mass spectrum of 4D gauge fields for $\theta_H = 0.1$ and $m_\KK = 13\,$TeV.
}
\vskip 10pt
\begin{tabular}{ccccccc}
\hline \hline 
$n$ &$\ell$  & $W^{(n)}$ & $Z^{(n)}$ &$\gamma^{(\ell)}$ &$W_R^{(\ell)}$, $Z_R^{(\ell)}$ &$A^{\hat 4 \, (\ell)}$\\
\hline
0 &0 &80.0\GeV &91.2\GeV &0 &$\cdots$ &$\cdots$\\
\hline
1 &1 &10.2\TeV &10.2\TeV &10.2\TeV &9.95\TeV &$\cdots$\\
\hline
2 & 1 &15.9\TeV &15.9\TeV &$\cdots$ &$\cdots$ &15.9\TeV\\
\hline
3 &2 &23.1\TeV &23.1\TeV &23.1TeV &22.8\TeV &$\cdots$\\
\hline
4 & 2 &29.0\TeV &29.0\TeV &$\cdots$ &$\cdots$ &29.0\TeV\\
\hline
5 & 3 &36.1\TeV &36.1\TeV &36.1TeV &35.8\TeV &$\cdots$\\
\hline
6& 3 &42.1\TeV &42.1\TeV &$\cdots$ &$\cdots$ &42.1\TeV\\
\hline \hline
\end{tabular}
\label{Table:mass}
\end{center}
\end{table}

\subsection{Wave functions at the UV and IR branes}

The anomaly coefficients for 4D gauge fields are proportional to the $F$ factors which are expressed 
in terms of various $\bar h$ functions defined in Eqs.\ (\ref{hbar1}) and (\ref{hbar2}) at the UV and IR branes.
The values of $\bar h^{L/R}_{W^{(\ell)}} (0)$, $\bar h^{L/R}_{W^{(\ell)}} (L)$
and $\bar h^{L/R}_{W_R^{(\ell)}} (L)$ are tabulated in Table \ref{Table:hbarfunc1}.
In the charged boson sector $\bar h^L_{W^{(0)}} (0), \bar h^L_{W^{(0)}} (L) \sim 1$.
$|\bar h^L_{W^{(2 \ell -1)}} (L) |$ and $| \bar h^{L/R}_{W_R^{(\ell)}} (L) |$ ($\ell = 1, 2, 3, \cdots$) are large.

\begin{table}[tbh]
\renewcommand{\arraystretch}{1.3}
\begin{center}
\caption{$\bar h^{L/R}_{W^{(\ell)}} (0)$, $\bar h^{L/R}_{W^{(\ell)}} (L)$
and $\bar h^{L/R}_{W_R^{(\ell)}} (L)$  in Eq.\ (\ref{hbar1}) for $\theta_H = 0.1$ and $m_\KK = 13\,$TeV
are tabulated.   Note that $\bar h^{L/R}_{W_R^{(\ell)}} (0) = 0$.
In the table $10^{-16}$ implies $O(10^{-16} )$.
}
\vskip 10pt
\begin{tabular}{crccccc}
\hline \hline 
$\ell$  & $\bar h^L_{W^{(\ell)}} (0)$ &  $\bar h^R_{W^{(\ell)}} (0)$ & $\bar h^L_{W^{(\ell)}} (L)$ & $\bar h^R_{W^{(\ell)}} (L)$ 
&$\bar h^L_{W_R^{(\ell)}} (L)$ &$\bar h^R_{W_R^{(\ell)}} (L)$ \\
\hline
0 &$1.0001$ &$10^{-16}$ &$0.9977$ &$0.0025$  &$\cdots$ &$\cdots$  \\
\hline
1 &$-0.2225$ &$10^{-17}$ &$7.3070$ &$0.0183$  &$0.0209$ &$-8.3314$ \\
\hline
2 &$-0.0132$ &$10^{-18}$ &$0.0328$ &$0.0001$  &$0.0209$ &$-8.3314$ \\
\hline
3 &$0.1536$ &$10^{-17}$ &$7.3052$ &$0.0183$   &$0.0209$ &$-8.3314$ \\
\hline
4 &$0.0099$ &$10^{-18}$ &$0.0329$ &$0.0001$   &$0.0209$ &$-8.3314$ \\
\hline
5&$-0.1252$ &$10^{-17}$ &$7.3046$ &$0.0183$   &$0.0209$ &$-8.3314$ \\
\hline \hline
\end{tabular}
\label{Table:hbarfunc1}
\end{center}
\end{table}

The values of $\bar h^{L/R, su2}_{Z^{(\ell)}} (0)$, $\bar h^{L/R, su2}_{Z^{(\ell)}} (L)$, $\bar h^{em}_{Z^{(\ell)}} (0)$,
$\bar h^{em}_{Z^{(\ell)}} (L)$, $\bar h_{\gamma^{(\ell)}} (0)$, $\bar h_{\gamma^{(\ell)}} (L)$ 
and $\bar h^{L/R}_{Z_R^{(\ell)}} (L)$
are tabulated in Tables \ref{Table:hbarfunc2} and  \ref{Table:hbarfunc3}.
In the neutral boson sector $\bar h_{\gamma^{(0)}} (0) = \bar h_{\gamma^{(0)}} (L) =1$ and
$\bar h^{L, su2}_{Z^{(0)}} (L), \bar h^{em}_{Z^{(0)}} (L) \sim 1$.
$|\bar h^{L, su2}_{Z^{(2 \ell -1)}} (L) |$, $| \bar h^{em}_{Z^{(2 \ell - 1)}} (L) |$,
$| \bar h_{\gamma^{(\ell)}} (L) |$ and $| \bar h^R_{Z_R^{(\ell)}} (L) |$  ($\ell = 1, 2, 3, \cdots$) are large.

\begin{table}[tbh]
\renewcommand{\arraystretch}{1.3}
\begin{center}
\caption{$\bar h^{L/R, su2}_{Z^{(\ell)}} (0)$, $\bar h^{L/R, su2}_{Z^{(\ell)}} (L)$, $\bar h^{em}_{Z^{(\ell)}} (0)$
and $\bar h^{em}_{Z^{(\ell)}} (L)$  in Eq.\  (\ref{hbar2}) for $\theta_H = 0.1$ and $m_\KK = 13\,$TeV
are tabulated.  Note that $\bar h^{em}_{Z^{(\ell)}} (y) = \bar h^{L, su2}_{Z^{(\ell)}} (y) + \bar h^{R, su2}_{Z^{(\ell)}} (y)$.
In the table $10^{-16}$ implies $O(10^{-16} )$.
}
\vskip 10pt
\begin{tabular}{crcccrc}
\hline \hline 
$\ell$  & $\bar h^{L, su2}_{Z^{(\ell)}} (0)$ &  $\bar h^{R, su2}_{Z^{(\ell)}} (0)$ &  $\bar h^{L, su2}_{Z^{(\ell)}} (L)$ 
&  $\bar h^{R, su2}_{Z^{(\ell)}} (L)$  &$\bar h^{em}_{Z^{(\ell)}} (0)$ &$\bar h^{em}_{Z^{(\ell)}} (L)$\\
\hline
0 &$1.0001$ &$10^{-16}$ &$0.9977$ &$0.0025$  &$1.0001$ &$1.0002$  \\
\hline
1 &$-0.2223$ &$10^{-17}$ &$7.3124$ &$0.0183$  &$-0.2223$ &$7.3307$ \\
\hline
2 &$-0.0151$ &$10^{-18}$ &$0.0374$ &$0.0001$  &$-0.0151$ &$0.0375$ \\
\hline
3 &$0.1535$ &$10^{-17}$ &$7.3107$ &$0.0183$   &$0.1535$ &$7.3290$ \\
\hline
4 &$0.0113$ &$10^{-18}$ &$0.0375$ &$0.0001$   &$0.0113$ &$0.0376$ \\
\hline
5&$-0.1252$ &$10^{-17}$ &$7.3101$ &$0.0183$   &$-0.1252$ &$7.3284$ \\
\hline \hline
\end{tabular}
\label{Table:hbarfunc2}
\end{center}
\end{table}

\begin{table}[tbh]
\renewcommand{\arraystretch}{1.3}
\begin{center}
\caption{$\bar h_{\gamma^{(\ell)}} (0)$, $\bar h_{\gamma^{(\ell)}} (L)$ and $\bar h^{L/R}_{Z_R^{(\ell)}} (L)$
in Eq.\  (\ref{hbar2}) for $\theta_H = 0.1$ and $m_\KK = 13\,$TeV are tabulated. 
Note that $\bar h^{L/R}_{Z_R^{(\ell)}} (0) =0$.
}
\vskip 10pt
\begin{tabular}{crccc}
\hline \hline 
$\ell$  & $\bar h_{\gamma^{(\ell)}} (0)$ &  $\bar h_{\gamma^{(\ell)}} (L)$ &  $\bar h^{L}_{Z_R^{(\ell)}} (L)$ 
&  $\bar h^{R}_{Z_R^{(\ell)}} (L)$  \\
\hline
0 &$1 ~~~$ &$1$ &$\cdots$ &$\cdots$  \\
\hline
1 &$-0.2230$ &$7.3071$ &$0.0134$ &$-5.3676$  \\
\hline
2 &$0.1540$ &$7.3053$ &$0.0134$ &$-5.3676$   \\
\hline
3 &$-0.1256$ &$7.3048$ &$0.0134$ &$-5.3676$  \\
\hline
4 &$0.1091$ &$7.3045$ &$0.0134$ &$-5.3676$  \\
\hline
5&$-0.0979$ &$7.3043$ &$0.0134$ &$-5.3676$   \\
\hline \hline
\end{tabular}
\label{Table:hbarfunc3}
\end{center}
\end{table}

\subsection{$F$ factors}

$F$ actors in Eqs.\ (\ref{ChargedAnom2}) and (\ref{NeutralAnom3}) are determined by the values of $\bar h (y)$
functions at $y=0$ and $L$.  Some of the $\bar h (L)$ for the KK excited modes of gauge bosons are large,
which give large $F$ factors and anomaly coefficients.  We list a few examples here.

$F_{W^{(\ell)} W^{\dagger (r)}}$ ($\ell, r = 0 \sim 5$) is given by
\begin{align}
F_{W^{(\ell)} W^{\dagger (r)}} &=
\begin{pmatrix}
0.9977	&3.5337	&0.0097	&3.7208	&0.0214	&3.5811  \cr
3.5337	&26.720	&0.1212	&26.672	&0.1191	&26.701  \cr
0.0097	&0.1212	&0.0006	&0.1187	&0.0005	&0.1205  \cr
3.7208	&26.672	&0.1187	&26.695	&0.1210	&26.671  \cr
0.0214	&0.1191	&0.0005	&0.1210	&0.0006	&0.1196  \cr
3.5811	&26.701	&0.1205	&26.671	&0.1196	&26.687
\end{pmatrix} .
\label{FWWtable}
\end{align}
$F^1_{Z^{(\ell)} Z^{(r)}}$ ($\ell, r = 0 \sim 5$) is given by
\begin{align}
F^1_{Z^{(\ell)} Z^{(r)}} &=
\begin{pmatrix}
0.9978	&3.5367	&0.0111	&3.7237	&0.0244	&3.5841  \cr
3.5367	&26.760	&0.1383	&26.712	&0.1359	&26.741  \cr
0.0111	&0.1383	&0.0008	&0.1354	&0.0006	&0.1375  \cr
3.7237	&26.712	&0.1354	&26.734	&0.1380	&26.711  \cr
0.0244	&0.1359	&0.0006	&0.1380	&0.0008	&0.1364  \cr
3.5841	&26.741	&0.1375	&26.711	&0.1364	&26.727
\end{pmatrix} .
\label{FZZtable}
\end{align}
$F^1_{Z^{(\ell)} \gamma^{(r)}}$ ($\ell, r = 0 \sim 5$) is given by
\begin{align}
F^1_{Z^{(\ell)} \gamma^{(r)}} &=
\begin{pmatrix}
0.9976	&3.5245	&3.7122	&3.5721	&3.6893	&3.5857  \cr
3.5359	&26.674	&26.626	&26.655	&26.628	&26.650  \cr
0.0111	&0.1378	&0.1349	&0.1370	&0.1353	&0.1368  \cr
3.7229	&26.626	&26.648	&26.625	&26.642	&26.625  \cr
0.0244	&0.1355	&0.1376	&0.1360	&0.1373	&0.1361  \cr
3.5833	&26.655	&26.625	&26.640	&26.625	&26.637
\end{pmatrix} .
\label{FZgammatable}
\end{align}
Notice that $F_{W^{(0)} W^{\dagger (0)}}, F^1_{Z^{(0)} Z^{(0)}}, F^1_{Z^{(0)} \gamma^{(0)}} \sim 1$, i.e.\ 
the deviation from the SM value ($=1$) is very small.

Other $F$ factors are found to be
\begin{align}
F_{W_R^{(\ell)} W_R^{\dagger (r)}} &\sim -34.7 ~~~ (\ell, r \ge 1) , \cr
\noalign{\kern 5pt}
F_{W_R^{(\ell)}W^{\dagger (r)}} &\sim 
\begin{cases} 0.02 &(\ell \ge 1, r =0) \cr  0.15 &(\ell \ge 1, r=1,3,5, \cdots) \cr 0.0007 & (\ell \ge 1, r=2,4,6, \cdots)  \end{cases} .\cr
\noalign{\kern 5pt}
F_{Z_R^{(\ell)} Z_R^{(r)}} &\sim -14.4  ~~~ (\ell, r \ge 1) , \cr
\noalign{\kern 5pt}
F_{Z_R^{(\ell)} \gamma^{(r)}} &\sim 
\begin{cases} 2.69 &(\ell \ge 1, r =0) \cr  19.7 &(\ell, r \ge 1) \end{cases} , \cr
\noalign{\kern 5pt}
F^1_{Z_R^{(\ell)}Z^{(r)}} &\sim 
\begin{cases} 0.01 &(\ell \ge 1, r =0) \cr  0.098 &(\ell \ge 1, r=1,3,5, \cdots) \cr 0.0005 & (\ell \ge 1, r=2,4,6, \cdots)  \end{cases} , \cr
\noalign{\kern 5pt}
F^2_{Z_R^{(\ell)}Z^{(r)}} &\sim 
\begin{cases} 2.69 &(\ell \ge 1, r =0) \cr  19.7 &(\ell \ge 1, r=1,3,5, \cdots) \cr 0.10 & (\ell \ge 1, r=2,4,6, \cdots)  \end{cases} .
\label{Fothers}
\end{align}

\subsection{Universality relations in $F$ factors}

The universality relations, (\ref{anomWW3}), (\ref{anomWWR2}), (\ref{anomZZ2}), (\ref{anomZgamma2}), 
(\ref{anomZRZR2}), (\ref{anogammaZR2}) and (\ref{anomZZR2}), are confirmed by the first method, 
namely by evaluating gauge couplings first and then doing KK sums.  
This analysis was done in ref.\ \cite{anomaly2026} for $W^{(0)}$, $Z^{(0)}$  and $\gamma^{(0)}$.
The result was tabulated in Tables IV -- XII there.  
The correspondence between the notation in the present paper and that in  \cite{anomaly2026} is
shown in Table \ref{Table:correspondence}.

\begin{table}[tbh]
\renewcommand{\arraystretch}{1.3}
\begin{center}
\caption{The correspondence between the notation in the present paper and that in  \cite{anomaly2026}  for the $F$ factors.
}
\vskip 10pt
\begin{tabular}{ccc}
\hline \hline 
Present paper  & In ref.\  \cite{anomaly2026}  \\
\hline
$F^{\alpha}_{W^{(0)} W^{\dagger (0)}}$  &$F^{\alpha}_{\gamma WW}$  &$\alpha = ud, tb$\\
\hline
$F^{1 \, \beta \beta}_{Z^{(0)} Z^{(0)}}$ &$ F^{\beta \, 1}_{\gamma ZZ} $ &$\beta = u, d, t, b$ \\
\hline
$F^{2 \, \beta \beta}_{Z^{(0)} Z^{(0)}}$ &$ F^{\beta \, 2}_{\gamma ZZ} $ &$\beta = u, d, t, b$ \\
\hline
$F^{1 \, \beta \beta}_{Z^{(0)} \gamma^{(0)}}$ &$ F^{\beta \, 1}_{Z^{(0)}} $ &$\beta = u, d, t, b$ \\
\hline \hline
\end{tabular}
\label{Table:correspondence}
\end{center}
\end{table}


\def\jnl#1#2#3#4{{#1}{\bf #2},  #3 (#4)}

\def\Zphys{{\em Z.\ Phys.} }
\def\jssc{{\em J.\ Solid State Chem.\ }}
\def\jpsJ{{\em J.\ Phys.\ Soc.\ Japan }}
\def\ptps{{\em Prog.\ Theoret.\ Phys.\ Suppl.\ }}
\def\PTP{{\em Prog.\ Theoret.\ Phys.\  }}
\def\PTEP{{\em Prog.\ Theoret.\ Exp.\  Phys.\  }}
\def\JMP{{\em J. Math.\ Phys.} }
\def\NPB{{\em Nucl.\ Phys.} B}
\def\NP{{\em Nucl.\ Phys.} }
\def\PLB{{\it Phys.\ Lett.} B}
\def\PL{{\em Phys.\ Lett.} }
\def\PRL{\em Phys.\ Rev.\ Lett. }
\def\PRB{{\em Phys.\ Rev.} B}
\def\PRD{{\em Phys.\ Rev.} D}
\def\PRe{{\em Phys.\ Rep.} }
\def\AP{{\em Ann.\ Phys.\ (N.Y.)} }
\def\RMP{{\em Rev.\ Mod.\ Phys.} }
\def\ZPC{{\em Z.\ Phys.} C}
\def\SCI{{\em Science} }
\def\CMP{\em Comm.\ Math.\ Phys. }
\def\MPLA{{\em Mod.\ Phys.\ Lett.} A}
\def\IJMPA{{\em Int.\ J.\ Mod.\ Phys.} A}
\def\IJMPB{{\em Int.\ J.\ Mod.\ Phys.} B}
\def\EPJC{{\em Eur.\ Phys.\ J.} C}
\def\EPJP{{\em Eur.\ Phys.\ J.} Plus}
\def\PR{{\em Phys.\ Rev.} }
\def\JHEP{{\em JHEP} }
\def\JCAP{{\em JCAP} }
\def\cmp{{\em Com.\ Math.\ Phys.}}
\def\JPA{{\em J.\  Phys.} A}
\def\JPG{{\em J.\  Phys.} G}
\def\NJP{{\em New.\ J.\  Phys.} }
\def\CQG{\em Class.\ Quant.\ Grav. }
\def\ATMP{{\em Adv.\ Theoret.\ Math.\ Phys.} }
\def\ibid{{\em ibid.} }
\def\ChP{{\em Chin.Phys.}C}
\def\NCA{{\it Nuovo Cim.} A}

\renewenvironment{thebibliography}[1]
         {\begin{list}{[$\,$\arabic{enumi}$\,$]}  
         {\usecounter{enumi}\setlength{\parsep}{0pt}
          \setlength{\itemsep}{0pt}  \renewcommand{\baselinestretch}{1.2}
          \settowidth
         {\labelwidth}{#1 ~ ~}\sloppy}}{\end{list}}

\newpage
 \vskip 1.cm

\leftline{\Large \bf References}


\end{document}